\documentclass[12pt]{article}

\global\arraycolsep=1pt

\usepackage{amsbsy,amssymb,latexsym,amsfonts, amsmath}
\usepackage{mathrsfs}
\usepackage{graphicx}

\numberwithin{equation}{section}
\newcommand{\bel}[1]{\begin{equation}\label{#1}}                     
\newcommand{\bal}[1]{\begin{eqnarray}\label{#1}}                     
\newcommand{\be}{\begin{equation}}
\newcommand{\ee}{\end{equation}}

\newcommand{\im}{\mathrm{i}}
\newcommand{\ex}{\mathrm{e}}
\newcommand{\de}{\mathrm{d}}
\newcommand{\dis}{\displaystyle}

\newcommand{\qq}{\qquad}

\renewcommand{\thefootnote}{\fnsymbol{footnote}}
\newcommand{\bea}{\begin{equation}}
\newcommand{\eea}{\end{equation}}

\begin{document}
%%%%%%%%%%%%%%%%%%%%%%%%%%%%%%%%%%%%%%%%%%%%%%%%%%%%%%%%%%%%%%%%%%%%%%%%%%%%%%%%%%%%%%%%%%
%
% title page
%
%%%%%%%%%%%%%%%%%%%%%%%%%%%%%%%%%%%%%%%%%%%%%%%%%%%%%%%%%%%%%%%%%%%%%%%%%%%%%%%%%%%%%%%%%
\begin{titlepage}
%%%%%%%%%%%%%%%%%%%% preprint # %%%%%%%%%%%%%%%%%%
\begin{flushright}
\normalsize
%\filename
~~~~
OCU-PHYS 333\\
August, 2010 \\
\end{flushright}
%%%%%%%%%%%%%%%%%%%%%%%%%%%%%%%%%%%%%%%%%%%%%%%%%%

\vspace{15pt}

%%%%%%%%%%%%%%%%%%%% title %%%%%%%%%%%%%%%%%%%%%%%
\begin{center}
{\LARGE
 Massive Scaling Limit of  } \\
\vspace{5pt}
{\LARGE  $\beta$-Deformed Matrix Model of Selberg Type  }\\
\end{center}
%%%%%%%%%%%%%%%%%%%%%%%%%%%%%%%%%%%%%%%%%%%%%%%%%%

\vspace{23pt}

%%%%%%%%%%%%%%%%%%% authors %%%%%%%%%%%%%%%%%%%%%%
\begin{center}
{ H. Itoyama$^{a, b} $\footnote{e-mail: itoyama@sci.osaka-cu.ac.jp},
T. Oota$^b$\footnote{e-mail: toota@sci.osaka-cu.ac.jp}
  and 
 N. Yonezawa$^b$\footnote{e-mail: yonezawa@sci.osaka-cu.ac.jp} 
}\\
%%%%%%%%%%%%%%%%%%%%%%%%%%%%%%%%%%%%%%%%%%%%%%%%%%
%
\vspace{18pt}
%
%%%%%%%%%%%%%%%%%%% affiliation %%%%%%%%%%%%%%%%%%%

$^a$ \it Department of Mathematics and Physics, Graduate School of Science\\
Osaka City University\\
\vspace{5pt}

$^b$ \it Osaka City University Advanced Mathematical Institute (OCAMI)

\vspace{5pt}

3-3-138, Sugimoto, Sumiyoshi-ku, Osaka, 558-8585, Japan \\

\end{center}
%%%%%%%%%%%%%%%%%%%%%%%%%%%%%%%%%%%%%%%%%%%%%%%%%%%
%
\vspace{20pt}
\begin{center}
Abstract\\
\end{center}
%%%%%%%%%%%%%%%%%%%% abstract %%%%%%%%%%%%%%%%%%%%%
We consider  a series of massive scaling limits
$m_1 \rightarrow \infty$, $q \rightarrow 0$,
$\lim m_1 q = \Lambda_{3}$ followed by $m_4 \rightarrow \infty$, $\Lambda_{3} \rightarrow 0$,
$\lim  m_4 \Lambda_{3} = (\Lambda_2)^2 $ 
of the $\beta$-deformed matrix model of Selberg type ($N_c=2$, $N_f=4$) which reduce the number of 
flavours to $N_f=3$ and subsequently to $N_f=2$. This keeps
 the other parameters of the model finite, which include
  $n=N_L$ and $N=n+N_R$, namely, the size of the matrix and the "filling fraction". 
Exploiting the method developed before,
we generate  instanton expansion with finite $g_s, \epsilon_{1,2}$ to check
 the Nekrasov coefficients ($N_f =3,2$ cases) to the lowest order. 
 The limiting expressions provide integral representation of irregular conformal blocks
   which contains a $2d$  operator $\lim 
\frac{1}{C(q)} : \ex^{(1/2) \alpha_1 \phi(0)}:
\left( \int_0^q \de z : \ex^{b_E \phi(z)}: \right)^n : \ex^{(1/2) \alpha_2 \phi(q)}:$
 and is subsequently analytically continued.  
   
%%%%%%%%%%%%%%%%%%%%%%%%%%%%%%%%%%%%%%%%%%%%%%%%%%%

\vfill

\setcounter{footnote}{0}
\renewcommand{\thefootnote}{\arabic{footnote}}

\end{titlepage}

%%%%%%%%%%%%%%%%%%%%
\renewcommand{\thefootnote}{\arabic{footnote}}
\setcounter{footnote}{0}
%%%%%%%%%%%%%%%%%%%%

%%%%%%%%%%%%%%%%%%%%%%%%%%%%%%%%%%%%%%%%%%%%%%%%%%%%%%%%%%%%%%%%%%%%%%%%%%%%%%%%%%%%%%%%%%%%%%
\section{Introduction}
%%%%%%%%%%%%%%%%%%%%%%

 There has already been an ample amount of literature on the Seiberg-Witten prepotential 
 for the cases when massive flavours are present.
Just restricting our attention  to  the $SU(2)$ case, 
a partial list includes \cite{SW9408,HO,APS,Oht1,Oht2,DHKP9609,MasSuz,DHKP9610,DKM9611}.
In particular, ref. \cite{Oht1,Oht2} compute the prepotentials of lower flavours as
decoupling limits of those of higher flavors.

In the recent intense activities on the conjectured equivalence
between the (irregular) conformal block and the Nekrasov partition function
\cite{AGT,Wyllard,MMM0907,MM0908a,MM0908b} (for a partial proof, \cite{FatLit,HJS}),
the discussions of these decoupling limits are 
further advanced and augmented to contain the parameters $g_s$,
$\epsilon_{1,2}$ of genus expansion and quantization and are derived
both from $2d-6d$ perspective of the Riemann surface \cite{gai0908}
and from the Shapovalov form \cite{MMM0909a,MMM0909b,AM0910}.

Somewhat separately, the relevance of the $\beta$ deformation
of the one-matrix model \cite{DV} and that of the more general quiver
matrix model \cite{DV,IMO} to the above equivalence have been noted.
At the planar level, both the spectral curve of the one-matrix model \cite{DV}
and that of the quiver matrix model \cite{IMO} are shown to be
isomorphic to the corresponding Seiberg-Witten curve written in the Witten-Gaiotto
form \cite{Witten,Gaiotto}.
(For a check of the  decoupling limits at the planar free energy, see \cite{EM}.)

The connection between the conformal block and the $\beta$ deformed matrix model
becomes firmer through the Dotsenko-Fateev integral integral representation
\cite{DF,MMS0911,MMS1001,IO5}. In particular, the first few Nekrasov coefficients are
 derived and the $0d-4d$ dictionary has been established in \cite{IO5}. 
See also \cite{BT0909,MMM1003,KPW,MS1004,AY1004,NX1005,Tai1006,NX1006,KMS1007}.
This representation permits rigorous treatments for arbitrary
values of $g_s$ and $\epsilon_{1,2}$. In this paper, we will consider a series of
 massive scaling limits   which reduce the number of 
flavours to $N_f=3$ and subsequently to $N_f=2$, using the technology established in \cite{IO5}.
 The way in which these limits are taken is different from that considered on the basis of
  the Shapovalov form \cite{MMM0909a,AM0910}. 
 The limiting expressions provide integral representation of irregular conformal blocks
   which contains a $2d$  operator $\lim 
\frac{1}{C(q)} : \ex^{(1/2) \alpha_1 \phi(0)}:
\left( \int_0^q \de z : \ex^{b_E \phi(z)}: \right)^n : \ex^{(1/2) \alpha_2 \phi(q)}:$
 and is subsequently analytically continued.    

A similar consideration at higher (say five) point conformal block yields 
interesting chiral $3$ and $4$ point functions to study.

In the next section, we briefly recall \cite{IO5}.
In the  section three,  we take the massive scaling limit to the three flavour case
  and generate its first expansion coefficient. 
In  section four,  we subsequently  take the limit to the two flavour case
 and generate its first expansion coefficient. 
In the appendix,  the contour deformations assumed in section three and four are justified. 

%%%%%%%%%%%%%%%%%%%%%%%%%%%%%%%%%%%%%%%%%%%%%%%%%%%%%%%%%%%%%%%%%%%%%%%%%%%%%%%%%%%%%%%%%%%%%%%%
\section{Review of generic four point conformal block represented by 
Selberg  type matrix model}
%%%%%%%%%%%%%%%%%%%%%%%%%%%

The Dotsenko-Fateev multiple integral is an integral representation
of the generic $4$-point conformal block 
$\mathcal{F}(q|c; \Delta_1, \Delta_2, \Delta_3, \Delta_4, \Delta_I)$.
In \cite{IO5}, we have managed to put this into the form of the perturbed 
double-Selberg matrix model. Renaming the same quantity as $\mathcal{F}$
as $Z_{\mathrm{pert}-(\mathrm{Selberg})^2}$, we have obtained
\bel{Z3P20}
\begin{split}
Z_{\mathrm{pert}-(\mathrm{Selberg})^2}&= q^{\sigma} (1 - q)^{(1/2)\alpha_2 \alpha_3} \cr
& \times \left(  \prod_{I=1}^{N_L} \int_0^1 \de x_I \right)
\prod_{I=1}^{N_L} x_I^{b_E \alpha_1} ( 1 - x_I)^{b_E \alpha_2}
(1 - q x_I )^{b_E \alpha_3}
\prod_{1 \leq I < J \leq N_L}
| x_I - x_J |^{2b_E^2} \cr
& \times
\left( \prod_{J=1}^{N_R} \int_0^1 \de y_J \right)
\prod_{J=1}^{N_R} y_J^{b_E \alpha_4}
(1 - y_J )^{b_E \alpha_3}
( 1 - q \, y_J )^{b_E \alpha_2}
\prod_{1 \leq I < J \leq N_R}
| y_I - y_J|^{2b_E^2} \cr
& \qq \qq \qq \qq \times
\prod_{I=1}^{N_L} \prod_{J=1}^{N_R} ( 1 - q\, x_I y_J )^{2b_E^2}.
\end{split}
\ee
Here $n := N_L$ and $N_R:= N-n$ are originally the number of the screening
operators we put between $0$ and $q$ and that between $1$ and $\infty$ respectively.
The remaining parameters are related to those of the original conformal block by
$c=1-6 Q_E^2$, $Q_E = b_E - (1/b_E)$, $\Delta_i = (1/4) \alpha_i( \alpha_i - 2 Q_E)$,
$\Delta_I = (1/4) \alpha_I ( \alpha_I - 2 Q_E)$. Also
\bel{constR}
\alpha_1 + \alpha_2 + \alpha_3 + \alpha_4 + 2 ( N_L + N_R) b_E = 2 Q_E.
\ee
\bel{sigma}
\sigma:= \frac{1}{2} \alpha_1 \alpha_2 + N_L
+  N_L b_E (\alpha_1 + \alpha_2) + N_L(N_L-1) b_E^2.
\ee

The Selberg integral is denoted by
\bel{SelI}
\begin{split}
& S_N(\beta_1, \beta_2, \gamma)
= \left( \prod_{I=1}^N \int_0^1 \de x_I \right)
\prod_{I=1}^{N} x_I^{\beta_1 -1} (1-x_I)^{\beta_2-1}
 \prod_{1 \leq I < J \leq N} | x_I - x_J |^{2\gamma}.
\end{split}
\ee
The Selberg integral \eqref{SelI} is convergent \cite{sel} and
equals to
\bel{SELB}
S_N(\beta_1, \beta_2, \gamma)
= \prod_{j=1}^N
\frac{
\Gamma(1 + j \gamma)
\Gamma(\beta_1 + (j-1) \gamma)
\Gamma(\beta_2 + (j-1) \gamma)}
{ \Gamma(1+\gamma) \Gamma(\beta_1 + \beta_2 + (N+j-2) \gamma)} ,
\ee
  when
 $N$ is a positive integer and  the complex parameters above obey
\be
\mathrm{Re}\, \beta_1 >0, \ \
\mathrm{Re}\, \beta_2 >0, \ \
\mathrm{Re}\, \gamma > - \mathrm{min}
\left\{
\frac{1}{N},
\frac{\mathrm{Re\, \beta_1}}{N-1},
\frac{\mathrm{Re\, \beta_2}}{N-1}
\right\}.
\ee

The perturbed double-Selberg model \eqref{Z3P20} 
has a well-defined $q$-expansion if
\bel{condalpha}
\mathrm{Re}(b_E \alpha_i) >-1, \qq
(i=1,2,3,4),
\ee
\be
\mathrm{Re}(b_E^2) > - \mathrm{min}
\left\{ \frac{1}{N_L}, \frac{1}{N_R},
\frac{\mathrm{Re}(b_E \alpha_1)+1}{N_L-1},
\frac{\mathrm{Re}(b_E \alpha_2)+1}{N_L-1},
\frac{\mathrm{Re}(b_E \alpha_3)+1}{N_R-1},
\frac{\mathrm{Re}(b_E \alpha_4)+1}{N_R-1}
\right\},
\ee
and $|q|<1$.

Let 
\be
\begin{split}
& Z_{(\mathrm{Selberg})^2}(b_E; N_L, \alpha_1, \alpha_2; N_R,
\alpha_4, \alpha_3) \cr
&= Z_{\mathrm{Selberg}}(b_E; N_L, \alpha_1, \alpha_2)
Z_{\mathrm{Selberg}}(b_E; N_R, \alpha_4, \alpha_3) \cr
&:= S_{N_L}(1 + b_E \alpha_1, 1 + b_E \alpha_2, b_E^2)\,
S_{N_R}(1 + b_E \alpha_4, 1 + b_E \alpha_3, b_E^2).
\end{split}
\ee

Averaging with respect to $Z_{(\mathrm{Selberg})^2}$,
$Z_{\mathrm{Selberg}}(N_L)$ and $Z_{\mathrm{Selberg}}(N_R)$
is denoted by
$\langle\!\langle \dotsm \rangle\!\rangle_{N_L,L_R}$,
$\langle\!\langle \dotsm \rangle\!\rangle_{N_L}$
and
$\langle\!\langle \dotsm \rangle\!\rangle_{N_R}$ respectively.

The $q$-expansion of the perturbed double-Selberg model
 is a special case of that of  more general perturbed Selberg model
  and is exactly calculable.
Consider the following  function
\bel{ZpS}
Z_{\mathrm{pert}-\mathrm{Selberg}}(\beta_1, \beta_2, \gamma; \{ g_i \}  )
:=S_N(\beta_1, \beta_2, \gamma)
\left\langle\!\!\!\left\langle
\exp\left( \sum_{I=1}^{N} W(x_I; g )\right)
\right\rangle\!\!\!\right\rangle_{N},
\ee
where the averaging is with respect to the Selberg integral \eqref{SelI}
and
\be
W(x; g )= \sum_{i=0}^{\infty} g_i x^i.
\ee
Let us expand the exponential of the potential
into the Jack polynomials 
\be
\exp\left( \sum_{I=1}^{N} W(x_I; \{ g_i \} )\right)
= \sum_{\lambda} C_{\lambda}^{(\gamma)}(g) \,
P_{\lambda}^{(1/\gamma)}(x).
\ee
Here $P_{\lambda}^{(1/\gamma)}(x)$ is a polynomial
of $x=(x_1, \dotsm, x_N)$ and  
$\lambda=(\lambda_1, \lambda_2, \dotsm)$ is a partition:
$\lambda_1 \geq \lambda_2 \geq \dotsm \geq 0$. Jack polynomials 
are the eigenstates of
\be
\sum_{I=1}^{N}
\left( x_I \frac{\partial}{\partial x_I} \right)^2
+ \gamma \sum_{1 \leq I <  J \leq N}
\left( \frac{x_I + x_J}{x_I - x_J} \right)
\left(
x_I \frac{\partial}{\partial x_I}
- x_J \frac{\partial}{\partial x_J}
\right),
\ee
with homogeneous degree $|\lambda|=\lambda_1+\lambda_2+
\dotsm$
and are normalized such that for dominance ordering
\be
P_{\lambda}^{(1/\gamma)}(x) = m_{\lambda}(x) + \sum_{\mu < \lambda}
a_{\lambda \mu} m_{\mu}(x).
\ee
Here $m_{\lambda}(x)$ is the monomial symmetric polynomial.

Let $\lambda'$ be the conjugate partition of $\lambda$, i.e., whose
diagram of partition is the transpose of that of $\lambda$ along the main diagonal.
Then Macdonald-Kadell integral \cite{mac1,kad,kan} implies that
\bel{MK}
\begin{split}
\Bigl\langle\!\!\Bigr\langle P_{\lambda}^{(1/\gamma)}(x)
\Bigr\rangle\!\! \Bigr\rangle_{N}
&= \prod_{i \geq 1}
\frac{\dis
\bigl(\, \beta_1+ (N-i) \gamma \, \bigr)_{\lambda_i} \
\bigl(\, (N + 1-i) \gamma \, \bigr)_{\lambda_i}}
{\dis
\bigl(\, \beta_1 + \beta_2 + (2 N-1 - i) \gamma \,
\bigr)_{\lambda_i}} \cr
& \times
\prod_{(i,j) \in \lambda}
\frac{1}{(\lambda_i - j + ( \lambda_j' -i +1 ) \gamma)},
\end{split}
\ee
where $(a)_n$ is the Pochhammer symbol:
\be
(a)_n = a (a+1) \dotsm (a+n-1), \qq
(a)_0 = 1.
\ee
Computation of \eqref{ZpS} reduces to that of the expansion coefficients $C_{\lambda}^{(\gamma)}(g)$.

An important relation established in \cite{IO5} is the $0d$-$4d$ version
of the AGT relation. It reads
 \begin{align} \label{0d4d}
b_E N_L &= \frac{a-m_2}{g_s},&
b_E N_R &= - \frac{a+m_3}{g_s}, \cr 
\alpha_1 &= \frac{1}{g_s}( m_2 - m_1 + \epsilon ),&
\alpha_2 &= \frac{1}{g_s}( m_2 + m_1 ), \cr
\alpha_3 &= \frac{1}{g_s}( m_3
 + m_4 ),&
\alpha_4 &= \frac{1}{g_s}( m_3 - m_4 + \epsilon ).
\end{align}
Also $b_E = \epsilon_1/g_s$ and $\epsilon= \epsilon_1 + \epsilon_2$,
$(1/b_E) = - \epsilon_2/g_s$.
These relations convert the seven parameters of the matrix model
\be
b_E, \ \  N_L, \ \ \alpha_1, \ \ \alpha_2, \ \ 
N_R, \ \ \alpha_4, \ \ \alpha_3
\ee
under the constraint \eqref{constR} 
into the six unconstrained parameters of the $\mathcal{N}=2$ $SU(2)$ gauge
theory with $N_f=4$:
\bel{GTP}
\frac{\epsilon_1}{g_s}, \ \ 
\frac{a}{g_s}, \ \ 
\frac{m_1}{g_s}, \ \ 
\frac{m_2}{g_s}, \ \ 
\frac{m_3}{g_s}, \ \
\frac{m_4}{g_s}.
\ee

%%%%%%%%%%%%%%%%%%%%%%%%%%%%%%%%%%%%%%%%%%%%%%%%%%%%%%%%%%%%%%%%%%%%%%%%%%%%%%%%%%%%%%%%%%%%%%%
\section{The limit $m_1 \rightarrow \infty$: from $N_f=4$ to $N_f=3$}
%%%%%%%%%%%%%%%%%%%%%%%%%%%%%%%%%%%%%%%

First, let us consider the limit $m_1 \rightarrow \infty$, $q \rightarrow 0$,
keeping $ \Lambda_3:= m_1 q$ finite. 
Due to the left-right reflection symmetry, this limit is equivalent to the limit 
$m_4 \rightarrow \infty$, $q \rightarrow 0$, with $m_4 q$ fixed.
Without loss of generality, we therefore restrict ourselves to the former one.
Let $q_3:= \Lambda_3/g_s$. Under this limit, the parameters 
$\alpha_3$, $\alpha_4$, $b_E$, $N_L$, $N_R$ are unchanged and
\be
\lim_{q \rightarrow 0} q \alpha_1 = - q_3, \qq
\lim_{q \rightarrow 0} q \alpha_2 = q_3.
\ee
Note that
\be
\alpha_1 + \alpha_2 = \frac{2m_2 + \epsilon}{g_s}
\ee
remains finite in this limit. 
This is why we take the limit $m_1 \rightarrow \infty$ instead of $m_2 \rightarrow \infty$.
But in the naive limit $m_1 \rightarrow \infty$ of $Z_{\mathrm{pert}-(\mathrm{Selberg})^2}$
\eqref{Z3P20} diverges since $\mathrm{Re}(b_E \alpha_1) \rightarrow - \infty$ which is in
the outside of the parameter region \eqref{condalpha}\footnote{For simplicity, we assume that
$b_E$ is real and positive.}. Hence, we should modify the multiple integral \eqref{Z3P20}
before taking the limit.
 
In order to examine this limit, we first rescale the integration variables $x_I$ as $z_I = q x_I$:
\be
\begin{split}
Z_{\mathrm{pert}-(\mathrm{Selberg})^2}
&= q^{(1/2) \alpha_1 \alpha_2} (1- q)^{(1/2)\alpha_2 \alpha_3}
\left( \prod_{I=1}^{N_L} \int_0^q \de z_I \right)
\left( \prod_{J=1}^{N_R} \int_0^1 \de y_J \right) \cr
& \qq \times
\prod_{I=1}^{N_L} z_I^{b_E (\alpha_1 + \alpha_2)}
\left( \frac{q}{z_I} - 1 \right)^{b_E \alpha_2} ( 1 - z_I )^{b_E \alpha_3}
\prod_{1 \leq I < J \leq N_L} | z_I - z_J |^{2b_E^2} \cr
& \qq \times
\prod_{J=1}^{N_R} y_J^{b_E \alpha_4 + 2 b_E^2 N_L}
( 1 - y_J)^{b_E \alpha_3} ( 1 - q y_J)^{b_E \alpha_2}
\prod_{1 \leq I < J \leq N_R} | y_I - y_J|^{2 b_E^2} \cr
& \qq \times \prod_{I=1}^{N_L} \prod_{J=1}^{N_R} \left( \frac{1}{y_J} - z_I \right)^{2 b_E^2}.
\end{split}
\ee
The multiple integral part can be written as follows
\be
\left( \prod_{J=1}^{N_R} \int_0^1 \de y_J \right) \Phi(y) 
\prod_{J=1}^{N_R} y_J^{b_E \alpha_4 + 2 b_E^2 N_L}
( 1 - y_J)^{b_E \alpha_3} ( 1 - q y_J)^{b_E \alpha_2}
\prod_{1 \leq I < J \leq N_R} | y_I - y_J|^{2 b_E^2},
\ee
where
\be
\begin{split}
\Phi(y) &:=  \left( \prod_{I=1}^{N_L} \int_0^q \de z_I \right)
\prod_{I=1}^{N_L} z_I^{b_E (\alpha_1 + \alpha_2)}
\left( \frac{q}{z_I} - 1 \right)^{b_E \alpha_2} ( 1 - z_I )^{b_E \alpha_3} \cr
& \qq  \times 
\prod_{1 \leq I < J \leq N_L} | z_I - z_J |^{2b_E^2}
\prod_{I=1}^{N_L} \prod_{J=1}^{N_R} \left( \frac{1}{y_J} - z_I \right)^{2 b_E^2}.
\end{split}
\ee
We assume that by using certain contour integral, the integration path $[0,q]$
can be converted to some path $\mathcal{C}'_q$\footnote{In the original variable $x_I$, this contour 
$\mathcal{C}'_q$ corresponds to the contour $\tilde{C}_{\rho}$ in the Appendix. See the left of Figure 3.}:
\be
\Phi(y) = C(q) \left( \prod_{I=1}^{N_L} \int_{\mathcal{C}'_q} \de z_I \right) \dotsm.
\ee
Here $C(q)$ is a constant which also depends on other parameters.
Justification of this assumption is given in Appendix.

In the limit $q\rightarrow 0$,
\be
\begin{split}
\lim_{q \rightarrow 0} \frac{1}{C(q)} \Phi(y)
& = \left( \prod_{I=1}^{N_L} \int_{\mathcal{C}'_0} d z_I \right)
\prod_{I=1}^{N_L} z_I^{b_E (\alpha_1 + \alpha_2)}
\exp\left( - \frac{b_E q_3}{z_I} \right) ( 1 - z_I )^{b_E \alpha_3} \cr
& \qq  \times 
\prod_{1 \leq I < J \leq N_L} ( z_I - z_J )^{2b_E^2}
\prod_{I=1}^{N_L} \prod_{J=1}^{N_R} \left( \frac{1}{y_J} - z_I \right)^{2 b_E^2}.
\end{split}
\ee
Therefore, we have
\be
\begin{split}
&Z^{(3)}:=\lim_{q \rightarrow 0} 
\frac{q^{-(1/2) \alpha_1 \alpha_2}(1-q)^{-(1/2)\alpha_2 \alpha_3}}{C(q)} 
Z_{\mathrm{pert}-(\mathrm{Selberg})^2} \cr
&= \left( \prod_{I=1}^{N_L} \int_{\mathcal{C}'_0} \de z_I \right)
\left( \prod_{J=1}^{N_R} \int_0^1 \de y_J \right) 
\prod_{I=1}^{N_L} z_I^{b_E (\alpha_1 + \alpha_2)}
\exp\left( - \frac{b_E q_3}{z_I} \right) ( 1 - z_I )^{b_E \alpha_3} 
\prod_{1 \leq I < J \leq N_L} ( z_I - z_J )^{2b_E^2} \cr
& \qq \times
\prod_{J=1}^{N_R} y_J^{b_E \alpha_4}
( 1 - y_J)^{b_E \alpha_3} \exp\left( - b_E q_3 \, y_J \right)
\prod_{1 \leq I < J \leq N_R} | y_I - y_J|^{2 b_E^2} 
\times \prod_{I=1}^{N_L} \prod_{J=1}^{N_R} \left( 1 - y_J z_I \right)^{2 b_E^2} \cr
&= ( b_E q_3 )^{\hat{\sigma}} \left( \prod_{I=1}^{N_L} \int_{\mathcal{C}} \de w_I \right)
\prod_{I=1}^{N_L} w_I^{b_E \hat{\alpha}_1}
\ex^{-w_I} \prod_{1 \leq I < J \leq N_L} ( w_I - w_J )^{2b_E^2} \cr
& \qq \times \left( \prod_{J=1}^{N_R} \int_0^1 \de y_J \right)
\prod_{J=1}^{N_R} y_J^{b_E \alpha_4} ( 1 - y_J)^{b_E \alpha_3}
\prod_{1 \leq I<J \leq N_R} | y_I - y_J |^{2 b_E^2} \cr
& \qq \times 
\prod_{I=1}^{N_L} \left( 1 - \frac{b_E q_3}{w_I} \right)^{b_E \alpha_3}
\prod_{J=1}^{N_R} \ex^{-b_E q_3 \, y_J}
\prod_{I=1}^{N_L} \prod_{J=1}^{N_R}
\left( 1 - \frac{b_E q_3 \, y_J}{w_I} \right)^{2b_E^2}.
\end{split}
\ee
Here we have changed $z_I = b_E q_3/w_I$ and
\be
b_E \hat{\alpha}_1:= -2-2(N_L-1)b_E^2 - b_E( \alpha_1 + \alpha_2),
\ee
\be
\hat{\sigma}:= N_L + b_E N_E ( \alpha_1 + \alpha_2) + N_L(N_L-1)b_E^2.
\ee
Using 0d-4d dictionary, we find
\be
\hat{\alpha}_1 = \frac{\epsilon - 2a}{g_s}.
\ee
The contour $\mathcal{C}$ for $w_I$ is shown in Fig. 1.

%%%%%%%%%%%%%%%%%%%%% figure 1 %%%%%%%%%%%%%%%%%%%%%%%%
\begin{figure}[ht]
\begin{center}
%WinTpicVersion3.08
\unitlength 0.1in
\begin{picture}( 48.5000, 16.0400)(  9.2000,-23.5400)
% CIRCLE 2 0 0 0
% 4 1725 1552 1749 1528 1839 1546 1839 1546
% 
\special{pn 8}%
\special{sh 0.600}%
\special{ar 1726 1552 34 34  0.0000000 6.2831853}%
% STR 2 0 3 0
% 3 1645 1632 1645 1692 5 0
% \Large{$0$}
\put(16.4500,-16.9200){\makebox(0,0){\Large{$0$}}}%
% VECTOR 2 0 3 0
% 2 925 1542 925 1608
% 
\special{pn 8}%
\special{pa 926 1542}%
\special{pa 926 1608}%
\special{fp}%
\special{sh 1}%
\special{pa 926 1608}%
\special{pa 946 1542}%
\special{pa 926 1556}%
\special{pa 906 1542}%
\special{pa 926 1608}%
\special{fp}%
% CIRCLE 2 0 3 0
% 4 1725 1552 2513 1405 2513 1405 2599 1696
% 
\special{pn 8}%
\special{ar 1726 1552 802 802  0.1632927 6.0987570}%
% VECTOR 2 0 3 0
% 2 5750 1470 2510 1400
% 
\special{pn 8}%
\special{pa 5750 1470}%
\special{pa 2510 1400}%
\special{fp}%
\special{sh 1}%
\special{pa 2510 1400}%
\special{pa 2576 1422}%
\special{pa 2564 1402}%
\special{pa 2578 1382}%
\special{pa 2510 1400}%
\special{fp}%
% VECTOR 2 0 3 0
% 2 2500 1690 5770 1640
% 
\special{pn 8}%
\special{pa 2500 1690}%
\special{pa 5770 1640}%
\special{fp}%
\special{sh 1}%
\special{pa 5770 1640}%
\special{pa 5704 1622}%
\special{pa 5718 1642}%
\special{pa 5704 1662}%
\special{pa 5770 1640}%
\special{fp}%
\end{picture}%
\end{center}
\caption{Integration contour $\mathcal{C}$.}
\label{figC}
\end{figure}
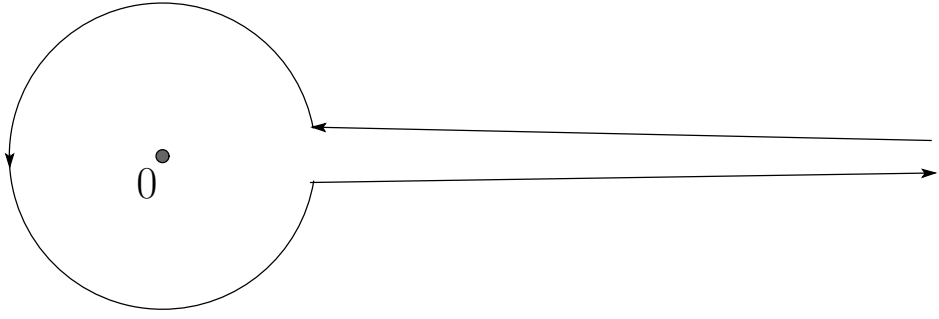
%%%%%%%%%%%%%%%%%%%%%%%%%%%%%%%%%%%%%%%%%%%%%%%%%%%%%%%%

The radius of the arc around the origin of this contour $C$ is assumed to be greater than $1$.  
Then, on the contour $\mathcal{C}$, 
it holds that $|w_I|>1$. Hence this multiple integral can serve as a well-defined generating function
of the $q_3$-expansion.

Without specifying the integration contours, the large $N$-limit of this type of ensemble average was 
studied in \cite{EM}.

Let
\be
T_N(\beta, \gamma):= \left( \prod_{I=1}^{N} \int_{\mathcal{C}}
d w_I \right) \prod_{I=1}^{N} w_I^{\beta-1} \ex^{-w_I} \prod_{1 \leq I<J \leq N}
( w_I - w_J )^{2\gamma}.
\ee
Now we have
\be
\begin{split}
Z^{(3)}&= (b_E q_3)^{\hat{\sigma}} T_{N_L}(1+b_E\hat{\alpha}_1, b_E^2)
S_{N_R}(1 + b_E \alpha_4, 1 + b_E \alpha_3, b_E^2) \cr
& \qq \times
\left\langle \!\!\! \left\langle
\prod_{I=1}^{N_L} \left( 1 - \frac{b_E q_3 }{w_I} \right)^{b_E \alpha_3}
\prod_{J=1}^{N_R} \ex^{-b_E q_3 \, y_J}
\prod_{I=1}^{N_L} \prod_{J=1}^{N_R}
\left( 1 - \frac{b_E q_3 \, y_J}{w_I} \right)^{2b_E^2}
\right\rangle\!\!\! \right\rangle_{N_L', N_R},
\end{split}
\ee
where the averaging $\langle\! \langle \dotsm \rangle\! \rangle_{N_L', N_R}$
is with respect to $T_{N_L}(1 + b_E \hat{\alpha}_1, b_E^2) S_{N_R}(1 + b_E \alpha_4,
1 + b_E \alpha_3, b_E^2)$.

Recall that $x_I = b_E q_3/(q w_I) = b_E m_1/(g_s w_I)$.
By taking the limit of the Macdonald-Kadell formula, we have
\be
\begin{split}
& \left\langle\!\!\left\langle P_{\lambda}^{(1/b_E^2)}(1/w) \right\rangle\!\!\right\rangle_{N_L'} \cr
&= \left( \frac{g_s^2}{\epsilon_1} \right)^{|\lambda|}
\prod_{(i,j) \in \lambda} 
\frac{(-a+m_2+\epsilon_1 (i-1) + \epsilon_2(j-1))}
{(2a-\epsilon - \epsilon_1 (i-1) - \epsilon_2 (j-1) )(\epsilon_1(\lambda'_j -i+1)
- \epsilon_2(\lambda_i-j))}.
\end{split}
\ee
Here $P_{\lambda}^{(1/b_E^2)}(1/w)$ is the Jack symmetric polynomials in $\{1/w_I\}_{1 \leq I \leq N_L}$
and $\lambda'$ is the conjugate partition of $\lambda$.

%%%%%%%%%%%%%%%%%%%%%%%%%%%%%%%%%%%%%%%%%%%%%%%%%%%%%%%%%%%%%%%%%%%%%%%%%%%%%%%%
\subsection{First expansion coefficient}
%%%%%%%%%%%%%%%%%%%%%%%%%%%%%%%%%%%%%%%%

Let us consider the following $\Lambda_3$-expansion:
\be
\begin{split}
\mathcal{A}^{(3)}(b_E q_3)
&:= \left\langle \!\!\! \left\langle
\prod_{I=1}^{N_L} \left( 1 - \frac{b_E q_3}{w_I} \right)^{b_E \alpha_3}
\prod_{J=1}^{N_R} \ex^{-b_E q_3 \, y_J}
\prod_{I=1}^{N_L} \prod_{J=1}^{N_R}
\left( 1 - \frac{b_E q_3 \, y_J}{w_I} \right)^{2b_E^2}
\right\rangle\!\!\! \right\rangle_{N_L', N_R} \cr
&= 1 + \sum_{\ell=1}^{\infty} \Lambda_3^{\ell} \mathcal{A}^{(3)}_{\ell}.
\end{split}
\ee
We have
\be
\mathcal{A}^{(3)}_1
= - \frac{\alpha_3}{g_s}  
\left\langle \!\!\! \left\langle
\sum_{I=1}^{N_L} \frac{b_E^2}{w_I} 
\right\rangle\!\!\! \right\rangle_{N_L'}
- \frac{1}{g_s} \left\langle\!\!\! \left\langle 
b_E \sum_{J=1}^{N_R} y_J 
\right\rangle \!\!\! \right\rangle_{N_R} 
- 2 \frac{1}{g_s} 
\left\langle \!\!\! \left\langle
\sum_{I=1}^{N_L} \frac{b_E^2}{w_I} 
\right\rangle\!\!\! \right\rangle_{N_L'}
\left\langle\!\!\! \left\langle 
b_E \sum_{J=1}^{N_R} y_J 
\right\rangle\!\!\! \right\rangle_{N_R}.
\ee
By using
\bel{LM1}
\left\langle\!\!\! \left\langle b_E^2 \sum_{I=1}^{N_L} \frac{1}{w_I} 
\right\rangle\!\!\!\right\rangle_{N_L'}
= \frac{b_E N_L}{\hat{\alpha}_1} = \frac{(a-m_2)}{(\epsilon - 2a)},
\ee
\be
\left\langle\!\!\!\left\langle b_E \sum_{J=1}^{N_R}
y_J \right\rangle\!\!\! \right\rangle_{N_R}
= \frac{b_E N_R(b_E N_R - Q_E + \alpha_4)}{(\alpha_3+\alpha_4 + 2 b_E N_R - 2 Q_E)}
= - \frac{(a+m_3)(a+m_4)}{g_s(2a+\epsilon)},
\ee
we have
\be
\mathcal{A}_1^{(3)}
= \frac{(a+m_2)(a+m_3)(a+m_4)}{2a(2a+\epsilon)g_s^2}
- \frac{(a-m_2)(a-m_3)(a-m_4)}{2a(2a-\epsilon)g_s^2}.
\ee
This is equivalent to the first Nekrasov function $Z_1^{\mathrm{Nek}}$ for $SU(2)$ with $N_f=3$.

%%%%%%%%%%%%%%%%%%%%%%%%%%%%%%%%%%%%%%%%%%%%%%%%%%%%%%%%%%%%%%%%%%%%%%%%%%%%%%%%%%%%%%%%%%%%%%
\section{The limit $m_4 \rightarrow \infty$: from $N_f=3$ to $N_f=2$}
%%%%%%%%%%%%%%%%%%%%%%%%%%%%%%%%%%%%%%%%%%%%%%%%%%%%%%%%%%%%%%%%%%%%%

Next, we consider the limit $m_4 \rightarrow \infty$, $\Lambda_3 \rightarrow 0$,
keeping $(\Lambda_2)^2:= m_4  \Lambda_3$ finite. Note that in this limit $q_3 = \Lambda_3/g_s \rightarrow 0$.
Let $q_2:= \Lambda_2/g_s$.
Under this limit, 
\be
\lim_{q_3 \rightarrow 0}  \alpha_3 q_3 = q_2^2, \qq
\lim_{q_3 \rightarrow 0}  \alpha_4 q_3 = - q_2^2,
\ee
and the following combination of the parameters
\be
\alpha_1 + \alpha_2 = \frac{2m_2+\epsilon}{g_s}, \qq
\alpha_3 + \alpha_4 = \frac{2m_3+\epsilon}{g_s},
\ee
remain finite and $b_E$, $N_L$, $N_R$
are unchanged.

Recall that
\be
\begin{split}
( b_E q_3)^{-\hat{\sigma}} Z^{(3)}
&= \left( \prod_{I=1}^{N_L} \int_{\mathcal{C}} \de w_I \right)
\prod_{I=1}^{N_L} w_I^{b_E \hat{\alpha}_1}
\ex^{-w_I} \prod_{1 \leq I < J \leq N_L} ( w_I - w_J )^{2b_E^2} \cr
& \qq \times \left( \prod_{J=1}^{N_R} \int_0^1 \de y_J \right)
\prod_{J=1}^{N_R} y_J^{b_E (\alpha_3+\alpha_4)} \left( \frac{1}{y_J} - 1 \right)^{b_E \alpha_3}
\prod_{1 \leq I<J \leq N_R} | y_I - y_J |^{2 b_E^2} \cr
& \qq \times 
\prod_{I=1}^{N_L} \left( 1 - \frac{b_E q_3}{w_I} \right)^{b_E \alpha_3}
\prod_{J=1}^{N_R} \ex^{-b_E q_3 \, y_J}
\prod_{I=1}^{N_L} \prod_{J=1}^{N_R}
\left( 1 - \frac{b_E q_3 \, y_J}{w_I} \right)^{2b_E^2}.
\end{split}
\ee
By setting $y_J = b_E \alpha_3/u_J$, we have
\be
\begin{split}
( b_E q_3)^{-\hat{\sigma}} Z^{(3)}
&= ( b_E \alpha_3)^{\hat{\sigma}'} 
\left( \prod_{I=1}^{N_L} \int_{\mathcal{C}} \de w_I \right)
\prod_{I=1}^{N_L} w_I^{b_E \hat{\alpha}_1}
\ex^{-w_I} \prod_{1 \leq I < J \leq N_L} ( w_I - w_J )^{2b_E^2} \cr
& \qq \times \left( \prod_{J=1}^{N_R} \int_{b_E \alpha_3}^{\infty} \de u_J \right)
\prod_{J=1}^{N_R} u_J^{b_E \hat{\alpha}_4} \left( \frac{u_J}{b_E \alpha_3} - 1 \right)^{b_E \alpha_3}
\prod_{1 \leq I<J \leq N_R} ( u_I - u_J )^{2 b_E^2} \cr
& \qq \times 
\prod_{I=1}^{N_L} \left( 1 - \frac{b_E q_3}{w_I} \right)^{b_E \alpha_3}
\prod_{J=1}^{N_R} \ex^{-b_E q_3 \, y_J}
\prod_{I=1}^{N_L} \prod_{J=1}^{N_R}
\left( 1 - \frac{b_E^2 q_3 \alpha_3}{w_I u_J} \right)^{2b_E^2},
\end{split}
\ee
where
\be
b_E \hat{\alpha}_4:= -2-b_E ( \alpha_3+\alpha_4) - 2 (N_R-1) b_E^2,
\ee
\be
\hat{\sigma}':= N_R + b_E N_R ( \alpha_3 + \alpha_4) + N_R (N_R-1) b_E^2.
\ee
Note that
\be
\hat{\alpha}_4 = \frac{\epsilon + 2a}{g_s}.
\ee
As in the first limit (from $N_f=4$ to $3$), we assume that the integration over $u_J$ in the path $[b_E \alpha_3, \infty]$
can be converted into contour integral over certain path $\mathcal{C}''_{q_3}$.
When converting all $u_J$ integrals, we denote an overall constant $C'(b_E q_3)$.

We also assume that
\be
\begin{split}
Z^{(2)}&:= \lim_{q_3 \rightarrow 0}
\frac{(b_E q_3)^{-\hat{\sigma}} (b_E \alpha_3)^{-\hat{\sigma}'}}{C'(b_E q_3)} Z^{(3)} \cr
&= \left( \prod_{I=1}^{N_L} \int_{\mathcal{C}} \de w_I \right)
\prod_{I=1}^{N_L} w_I^{b_E \hat{\alpha}_1}
\ex^{-w_I} \prod_{1 \leq I < J \leq N_L} ( w_I - w_J )^{2b_E^2} \cr
& \qq \times
\left( \prod_{J=1}^{N_R} \int_{\mathcal{C}} \de u_J \right)
\prod_{J=1}^{N_R} u_J^{b_E \hat{\alpha}_4} \ex^{-u_J}
\prod_{1 \leq I<J \leq N_R} ( u_I - u_J )^{2b_E^2} \cr
& \qq \times
\prod_{I=1}^{N_L} \exp\left( - \frac{(b_E q_2)^2}{w_I} \right)
\prod_{J=1}^{N_R} \exp\left( - \frac{(b_E q_2)^2}{u_J} \right)
\prod_{I=1}^{N_L} \prod_{J=1}^{N_R} \left(
1 - \frac{(b_E q_2)^2}{w_I u_J } \right)^{2b_E^2}.
\end{split}
\ee
Hence
\be
\begin{split}
Z^{(2)} &= T_{N_L}( 1 + b_E \hat{\alpha}_1, b_E^2) T_{N_R}( 1 + b_E \hat{\alpha}_4, b_E^2) \cr
& \qq \times
\left\langle\!\!\!\left\langle
\prod_{I=1}^{N_L} \exp\left( - \frac{(b_E q_2)^2}{w_I} \right)
\prod_{J=1}^{N_R} \exp\left( - \frac{(b_E q_2)^2}{u_J} \right)
\prod_{I=1}^{N_L} \prod_{J=1}^{N_R} \left(
1 - \frac{(b_E q_2)^2}{w_I u_J } \right)^{2b_E^2}
\right\rangle\!\!\!\right\rangle_{N_L', N_R'}.
\end{split}
\ee
Here the averaging $\langle\!\langle \dotsm \rangle\!\rangle_{N_L',N_R'}$
is with respect to $T_{N_L}( 1 + b_E \hat{\alpha}_1, b_E^2) T_{N_R}( 1 + b_E \hat{\alpha}_4, b_E^2)$.

Also we have the formula for the average of the Jack symmetric polynomials $P_{\lambda}^{(1/b_E^2)}(1/u)$
of $\{ 1/u_J \}_{1 \leq J \leq N_R}$:
\be
\begin{split}
& \left\langle\!\!\left\langle P_{\lambda}^{(1/b_E^2)}(1/u) \right\rangle\!\!\right\rangle_{N_R'} \cr
&= \left( - \frac{g_s^2}{\epsilon_1} \right)^{|\lambda|}
\prod_{(i,j) \in \lambda} 
\frac{(a+m_3+\epsilon_1 (i-1) + \epsilon_2(j-1))}
{(2a+\epsilon + \epsilon_1 (i-1) + \epsilon_2 (j-1) )(\epsilon_1(\lambda'_j -i+1)
- \epsilon_2(\lambda_i-j))}.
\end{split}
\ee

%%%%%%%%%%%%%%%%%%%%%%%%%%%%%%%%%%%%%%%%%%%%%%%%%%%%%%%%%%%%%%%%%%%%%%%%%%%%%%%%%%%%%%%%%%%%%%%
\subsection{First expansion coefficient}
%%%%%%%%%%%%%%%%%%%%%%%%%%%%%%%%%%%%%%%%

Let us consider the following $\Lambda_2$-expansion:
\be
\begin{split}
\mathcal{A}^{(2)}(b_E q_2)
&:=\left\langle\!\!\!\left\langle
\prod_{I=1}^{N_L} \exp\left( - \frac{(b_E q_2)^2}{w_I} \right)
\prod_{J=1}^{N_R} \exp\left( - \frac{(b_E q_2)^2}{u_J} \right)
\prod_{I=1}^{N_L} \prod_{J=1}^{N_R} \left(
1 - \frac{(b_E q_2)^2}{w_I u_J } \right)^{2b_E^2}
\right\rangle\!\!\!\right\rangle_{N_L', N_R'} \cr
&= 1 + \sum_{\ell=1}^{\infty} \Lambda_2^{2\ell} \mathcal{A}_{\ell}^{(2)}.
\end{split}
\ee
We have
\be
\mathcal{A}^{(2)}_1
= - \frac{1}{g_s^2}
\left\langle \!\!\! \left\langle
\sum_{I=1}^{N_L} \frac{b_E^2}{w_I} 
\right\rangle\!\!\! \right\rangle_{N_L'}
- \frac{1}{g_s^2} \left\langle\!\!\! \left\langle 
\sum_{J=1}^{N_R} \frac{b_E^2}{u_J}
\right\rangle \!\!\! \right\rangle_{N_R'} 
- \frac{2}{g_s^2} 
\left\langle \!\!\! \left\langle
\sum_{I=1}^{N_L} \frac{b_E^2}{w_I} 
\right\rangle\!\!\! \right\rangle_{N_L'}
\left\langle\!\!\! \left\langle 
\sum_{J=1}^{N_R} \frac{b_E^2}{u_J} 
\right\rangle\!\!\! \right\rangle_{N_R'}.
\ee
By using \eqref{LM1} and
\be
\left\langle\!\!\! \left\langle 
\sum_{J=1}^{N_R} \frac{b_E^2}{u_J}
\right\rangle \!\!\! \right\rangle_{N_R'} 
= \frac{b_E N_R}{\hat{\alpha}_4}
= - \frac{(a+m_3)}{(\epsilon+2a)},
\ee
we find
\be
\mathcal{A}_1^{(2)}
= \frac{(a+m_2)(a+m_3)}{2a(2a+\epsilon) g_s^2}
+ \frac{(a-m_2)(a-m_3)}{2a(2a-\epsilon) g_s^2},
\ee
which is the first Nekrasov function for $SU(2)$ with $N_f=2$.

%%%%%%%%%%%%%%%%%%%%%%%%%%%%%%%%%%%%%%%%%%%%%%%%%%%%%%%%%%%%%%%%%  acknowledgements
\section*{Acknowledgements}
We thank  Hiroaki Kanno for interesting discussions on this subject.
The research of H.~I.~ and T.~O.~
is supported in part by the Grant-in-Aid for Scientific Research (2054278)
  as well as JSPS Bilateral Joint Projects(JSPS-RFBR collaboration)
from the Ministry of Education, Science and Culture, Japan.

%%%%%%%%%%%%%%%%%%%%%%%%%%%%%%%%%%%%%%%%%%%%%%%%%%%%%%%%%%%%%%%%%%%%%%%%%%%%%%%%%%%%%%%%%%%%%%%%%%%%%

%%%%%%%%%%%%%%%%%%%%%%%%%%%%%%%%%%%%%%%%%%%%%%%%%%%%%%%%%%%%%%%%%%%%%

\appendix

%%%%%%%%%%%%%%%%%%%%%%%%%%%%%%%%%%%%%%%%%%%%%%%%%%%%%%%%%%%%%%%%%%%%%%%%%%%%%%
\section{Generalization of Integral representation of Iguri}
%%%%%%%%%%%%%%%%%%%%%%%%%%%%%%%%%%%%%%%%%%%%%%%%%%%%%%%%%%%%%%

In this appendix, 
    we obtain the analytic continuation of the Selberg average as multiple complex contour integrals
        where we can take our limits.
We employ a method similar to that used in \cite{I}.

\begin{figure}[tb]
    \begin{center}
        \begin{minipage}[t]{0.49\textwidth}
            \begin{center}
                \input{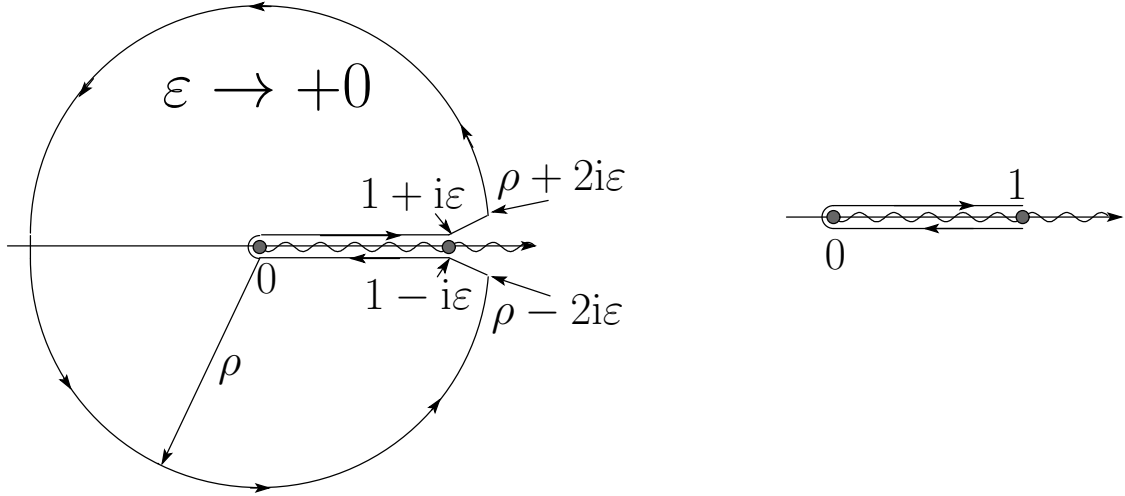}
            \end{center}
        \end{minipage}
        \hfill
        \begin{minipage}[t]{0.49\textwidth}
            \begin{center}
                %WinTpicVersion3.08
\unitlength 0.1in
\begin{picture}( 19.9000,  3.7700)(  0.8500, -17.2200)
%\begin{picture}( 19.9000,  3.7700)(  0.8500, -4.2200)
% FUNC 2 0 3 0
% 9 569 182 2009 422 569 302 857 302 569 542 569 182 2009 422 0 3 0 0
% 0.1*sin(10*x)
\special{pn 8}%
\special{pa 566 300}%
\special{pa 570 304}%
\special{pa 576 308}%
\special{pa 580 312}%
\special{pa 586 316}%
\special{pa 590 318}%
\special{pa 596 322}%
\special{pa 600 324}%
\special{pa 606 326}%
\special{pa 610 326}%
\special{pa 616 326}%
\special{pa 620 326}%
\special{pa 626 324}%
\special{pa 630 322}%
\special{pa 636 320}%
\special{pa 640 318}%
\special{pa 646 314}%
\special{pa 650 310}%
\special{pa 656 306}%
\special{pa 660 302}%
\special{pa 666 298}%
\special{pa 670 294}%
\special{pa 676 290}%
\special{pa 680 286}%
\special{pa 686 284}%
\special{pa 690 282}%
\special{pa 696 280}%
\special{pa 700 278}%
\special{pa 706 278}%
\special{pa 710 278}%
\special{pa 716 280}%
\special{pa 720 282}%
\special{pa 726 284}%
\special{pa 730 288}%
\special{pa 736 290}%
\special{pa 740 294}%
\special{pa 746 298}%
\special{pa 750 302}%
\special{pa 756 306}%
\special{pa 760 310}%
\special{pa 766 314}%
\special{pa 770 318}%
\special{pa 776 320}%
\special{pa 780 324}%
\special{pa 786 326}%
\special{pa 790 326}%
\special{pa 796 326}%
\special{pa 800 326}%
\special{pa 806 326}%
\special{pa 810 324}%
\special{pa 816 322}%
\special{pa 820 318}%
\special{pa 826 314}%
\special{pa 830 312}%
\special{pa 836 308}%
\special{pa 840 302}%
\special{pa 846 298}%
\special{pa 850 294}%
\special{pa 856 290}%
\special{pa 860 288}%
\special{pa 866 284}%
\special{pa 870 282}%
\special{pa 876 280}%
\special{pa 880 278}%
\special{pa 886 278}%
\special{pa 890 278}%
\special{pa 896 280}%
\special{pa 900 282}%
\special{pa 906 284}%
\special{pa 910 286}%
\special{pa 916 290}%
\special{pa 920 294}%
\special{pa 926 298}%
\special{pa 930 302}%
\special{pa 936 306}%
\special{pa 940 310}%
\special{pa 946 314}%
\special{pa 950 318}%
\special{pa 956 320}%
\special{pa 960 322}%
\special{pa 966 324}%
\special{pa 970 326}%
\special{pa 976 326}%
\special{pa 980 326}%
\special{pa 986 326}%
\special{pa 990 324}%
\special{pa 996 322}%
\special{pa 1000 318}%
\special{pa 1006 316}%
\special{pa 1010 312}%
\special{pa 1016 308}%
\special{pa 1020 304}%
\special{pa 1026 300}%
\special{pa 1030 296}%
\special{pa 1036 292}%
\special{pa 1040 288}%
\special{pa 1046 284}%
\special{pa 1050 282}%
\special{pa 1056 280}%
\special{pa 1060 280}%
\special{pa 1066 278}%
\special{pa 1070 278}%
\special{pa 1076 280}%
\special{pa 1080 282}%
\special{pa 1086 284}%
\special{pa 1090 286}%
\special{pa 1096 290}%
\special{pa 1100 292}%
\special{pa 1106 296}%
\special{pa 1110 300}%
\special{pa 1116 306}%
\special{pa 1120 310}%
\special{pa 1126 314}%
\special{pa 1130 316}%
\special{pa 1136 320}%
\special{pa 1140 322}%
\special{pa 1146 324}%
\special{pa 1150 326}%
\special{pa 1156 326}%
\special{pa 1160 326}%
\special{pa 1166 326}%
\special{pa 1170 324}%
\special{pa 1176 322}%
\special{pa 1180 320}%
\special{pa 1186 316}%
\special{pa 1190 312}%
\special{pa 1196 308}%
\special{pa 1200 304}%
\special{pa 1206 300}%
\special{pa 1210 296}%
\special{pa 1216 292}%
\special{pa 1220 288}%
\special{pa 1226 286}%
\special{pa 1230 282}%
\special{pa 1236 280}%
\special{pa 1240 280}%
\special{pa 1246 278}%
\special{pa 1250 278}%
\special{pa 1256 280}%
\special{pa 1260 280}%
\special{pa 1266 282}%
\special{pa 1270 286}%
\special{pa 1276 288}%
\special{pa 1280 292}%
\special{pa 1286 296}%
\special{pa 1290 300}%
\special{pa 1296 304}%
\special{pa 1300 308}%
\special{pa 1306 312}%
\special{pa 1310 316}%
\special{pa 1316 320}%
\special{pa 1320 322}%
\special{pa 1326 324}%
\special{pa 1330 326}%
\special{pa 1336 326}%
\special{pa 1340 326}%
\special{pa 1346 326}%
\special{pa 1350 324}%
\special{pa 1356 322}%
\special{pa 1360 320}%
\special{pa 1366 316}%
\special{pa 1370 314}%
\special{pa 1376 310}%
\special{pa 1380 306}%
\special{pa 1386 302}%
\special{pa 1390 296}%
\special{pa 1396 294}%
\special{pa 1400 290}%
\special{pa 1406 286}%
\special{pa 1410 284}%
\special{pa 1416 282}%
\special{pa 1420 280}%
\special{pa 1426 278}%
\special{pa 1430 278}%
\special{pa 1436 280}%
\special{pa 1440 280}%
\special{pa 1446 282}%
\special{pa 1450 284}%
\special{pa 1456 288}%
\special{pa 1460 292}%
\special{pa 1466 296}%
\special{pa 1470 300}%
\special{pa 1476 304}%
\special{pa 1480 308}%
\special{pa 1486 312}%
\special{pa 1490 316}%
\special{pa 1496 318}%
\special{pa 1500 322}%
\special{pa 1506 324}%
\special{pa 1510 326}%
\special{pa 1516 326}%
\special{pa 1520 326}%
\special{pa 1526 326}%
\special{pa 1530 324}%
\special{pa 1536 322}%
\special{pa 1540 320}%
\special{pa 1546 318}%
\special{pa 1550 314}%
\special{pa 1556 310}%
\special{pa 1560 306}%
\special{pa 1566 302}%
\special{pa 1570 298}%
\special{pa 1576 294}%
\special{pa 1580 290}%
\special{pa 1586 286}%
\special{pa 1590 284}%
\special{pa 1596 282}%
\special{pa 1600 280}%
\special{pa 1606 278}%
\special{pa 1610 278}%
\special{pa 1616 278}%
\special{pa 1620 280}%
\special{pa 1626 282}%
\special{pa 1630 284}%
\special{pa 1636 288}%
\special{pa 1640 290}%
\special{pa 1646 294}%
\special{pa 1650 298}%
\special{pa 1656 302}%
\special{pa 1660 306}%
\special{pa 1666 310}%
\special{pa 1670 314}%
\special{pa 1676 318}%
\special{pa 1680 320}%
\special{pa 1686 324}%
\special{pa 1690 326}%
\special{pa 1696 326}%
\special{pa 1700 326}%
\special{pa 1706 326}%
\special{pa 1710 326}%
\special{pa 1716 324}%
\special{pa 1720 320}%
\special{pa 1726 318}%
\special{pa 1730 314}%
\special{pa 1736 310}%
\special{pa 1740 306}%
\special{pa 1746 302}%
\special{pa 1750 298}%
\special{pa 1756 294}%
\special{pa 1760 290}%
\special{pa 1766 288}%
\special{pa 1770 284}%
\special{pa 1776 282}%
\special{pa 1780 280}%
\special{pa 1786 278}%
\special{pa 1790 278}%
\special{pa 1796 278}%
\special{pa 1800 280}%
\special{pa 1806 282}%
\special{pa 1810 284}%
\special{pa 1816 286}%
\special{pa 1820 290}%
\special{pa 1826 294}%
\special{pa 1830 298}%
\special{pa 1836 302}%
\special{pa 1840 306}%
\special{pa 1846 310}%
\special{pa 1850 314}%
\special{pa 1856 318}%
\special{pa 1860 320}%
\special{pa 1866 322}%
\special{pa 1870 324}%
\special{pa 1876 326}%
\special{pa 1880 326}%
\special{pa 1886 326}%
\special{pa 1890 326}%
\special{pa 1896 324}%
\special{pa 1900 322}%
\special{pa 1906 318}%
\special{pa 1910 316}%
\special{pa 1916 312}%
\special{pa 1920 308}%
\special{pa 1926 304}%
\special{pa 1930 300}%
\special{pa 1936 296}%
\special{pa 1940 292}%
\special{pa 1946 288}%
\special{pa 1950 284}%
\special{pa 1956 282}%
\special{pa 1960 280}%
\special{pa 1966 280}%
\special{pa 1970 278}%
\special{pa 1976 278}%
\special{pa 1980 280}%
\special{pa 1986 282}%
\special{pa 1990 284}%
\special{pa 1996 286}%
\special{pa 2000 290}%
\special{pa 2006 294}%
\special{sp}%
% VECTOR 2 0 3 0
% 2 323 302 2075 302
% 
\special{pn 8}%
\special{pa 324 302}%
\special{pa 2076 302}%
\special{fp}%
\special{sh 1}%
\special{pa 2076 302}%
\special{pa 2008 282}%
\special{pa 2022 302}%
\special{pa 2008 322}%
\special{pa 2076 302}%
\special{fp}%
% LINE 2 0 3 0
% 2 569 362 1559 362
% 
\special{pn 8}%
\special{pa 570 362}%
\special{pa 1560 362}%
\special{fp}%
% CIRCLE 2 0 0 0
% 4 569 302 593 278 683 296 683 296
% 
\special{pn 8}%
\special{sh 0.600}%
\special{ar 570 302 34 34  0.0000000 6.2831853}%
% CIRCLE 2 0 0 0
% 4 1559 302 1583 278 1673 296 1673 296
% 
\special{pn 8}%
\special{sh 0.600}%
\special{ar 1560 302 34 34  0.0000000 6.2831853}%
% STR 2 0 3 0
% 3 580 440 580 500 5 0
% \Large{$0$}
\put(5.8000,-5.0000){\makebox(0,0){\Large{$0$}}}%
% STR 2 0 3 0
% 3 1530 70 1530 130 5 0
% \Large{$1$}
\put(15.3000,-1.3000){\makebox(0,0){\Large{$1$}}}%
% VECTOR 2 0 3 0
% 2 1301 362 1049 362
% 
\special{pn 8}%
\special{pa 1302 362}%
\special{pa 1050 362}%
\special{fp}%
\special{sh 1}%
\special{pa 1050 362}%
\special{pa 1116 382}%
\special{pa 1102 362}%
\special{pa 1116 342}%
\special{pa 1050 362}%
\special{fp}%
% CIRCLE 2 0 3 0
% 4 569 302 563 362 569 -1030 569 1790
% 
\special{pn 8}%
\special{ar 570 302 60 60  1.5707963 4.7123890}%
% LINE 2 0 3 0
% 2 569 242 1559 242
% 
\special{pn 8}%
\special{pa 570 242}%
\special{pa 1560 242}%
\special{fp}%
% VECTOR 2 0 3 0
% 2 887 242 1301 242
% 
\special{pn 8}%
\special{pa 888 242}%
\special{pa 1302 242}%
\special{fp}%
\special{sh 1}%
\special{pa 1302 242}%
\special{pa 1234 222}%
\special{pa 1248 242}%
\special{pa 1234 262}%
\special{pa 1302 242}%
\special{fp}%
\end{picture}%
            \end{center}
        \end{minipage}
                \caption{{\small
                    $x_{n+1}$ plane is illustrated. 
                    The wiggly lines are cuts of $\Phi_N^{(0)}$ with $N=3$.
                    The left figure shows a contour $\mathcal{C}_{\rho}$.
                    Note that this contour does not get across the cuts of $\Phi_N^{(0)}$.
                    We denote by $\mathcal{C}_{[0,1]}$ the path appearing in the right figure.
                }}
                \label{fig:C_R+C_01}
    \end{center}
\end{figure}

\begin{figure}[tb]
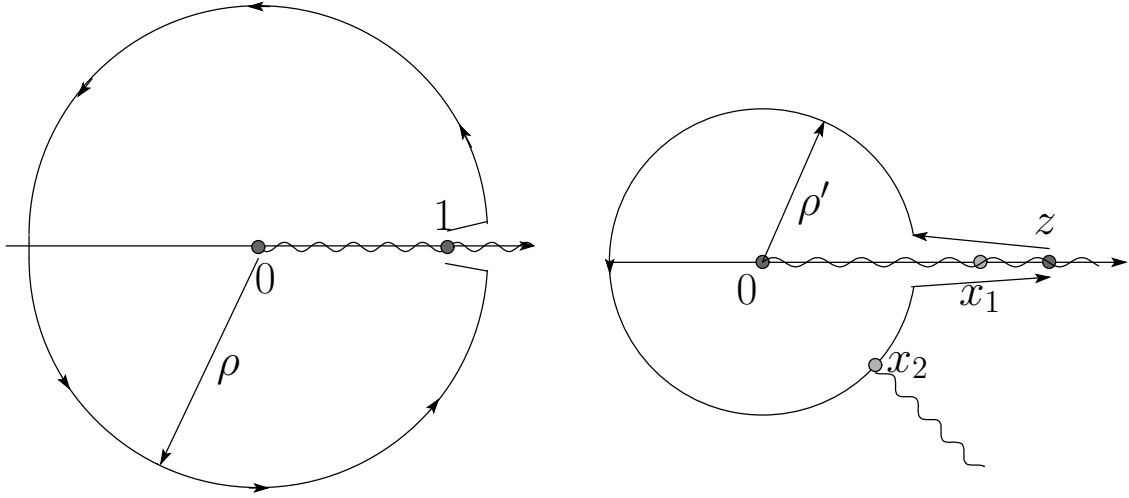

    \begin{center}
        \begin{minipage}[t]{0.49\textwidth}
            \begin{center}
                \input{fig3a.tex}
            \end{center}
        \end{minipage}
        \hfill
        \begin{minipage}[t]{0.49\textwidth}
            \begin{center}
                \input{fig3b.tex}
            \end{center}
        \end{minipage}
        \caption{{\small
            $x_{n+1}$ plane is illustrated. 
            The wiggly lines are cuts of $\Phi_N^{(0)}$ with $N=3$.
            A path $\tilde{\mathcal{C}}_{\rho}$ is illustrated in the left figure.
            This path is obtained by $\mathcal{C}_\rho-\mathcal{C}_{[0,1]}$.
            A path $\tilde{\mathcal{D}}_{\rho}$ is illustrated in the right figure.
            As a matter of convenience, we introduce parameter $z$.
            $z + 0 \im$ and $z - 0 \im$ are the starting point and the end point of $\tilde{\mathcal{D}}(\rho',z)$ respectively.
                }}
                \label{fig:CD}
    \end{center}
\end{figure}

Firstly, we introduce some definitions.
In this appendix, we study complex integrals along four paths specified at Fig.\ref{fig:C_R+C_01} and Fig.\ref{fig:CD}.
The symbol $\mathcal{C}_{\rho}$ denotes the contour shown in Fig.\ref{fig:C_R+C_01} (left).
The symbol $\mathcal{C}_{[0,1]}$ stands for a path composed of the segments connecting $x=0\pm 0\im$ and $x=1\pm 0\im$.
See Fig.\ref{fig:C_R+C_01} (right).
The symbol $\tilde{\mathcal{C}}_{\rho}$ is defined by subtracting of $\mathcal{C}_{[0,1]}$ from $\mathcal{C}_\rho$:
    $\tilde{\mathcal{C}}_{\rho}:=\mathcal{C}_{\rho}-\mathcal{C}_{[0,1]}$.
See Fig.\ref{fig:CD} (left).
The symbol $\tilde{\mathcal{D}}_{\rho}$ stands for a tadpole-type path specified at Fig.\ref{fig:CD} (right).
Let us introduce a symbol representing an ordering along the path  $\mathcal{C}_{\rho}$ by $\preceq$.
We mean by $x\preceq y$ that $y$ is ahead of $x$ along the path.
To complete the definition of $\preceq$, we regard $1-0-0\im$ and $1+0-0\im$ as the starting point and the end point of $\mathcal{C}_{\rho}$ respectively.
The symbol $\Phi_N^{(n)}$ represents an integral kernel used in this appendix: 
\begin{align}
\begin{split}
\Phi_N^{(n)}(\beta_1,\beta_2,\gamma;x):&=
\left(\prod_{I=1}^{n}|x_I|^{\beta_1-1} | 1 - x_I|^{\beta_2-1}\right)
\left(\prod_{\bar{I}=n+1}^{N}(-x_{\bar{I}})^{\beta_1-1} ( 1 - x_{\bar{I}})^{\beta_2-1}\right)\\
&\hspace{10mm}\times
\left(\prod_{1 \leq \hat{I} \leq n <\hat{J} \leq N}
    \left( 1-\frac{x_{\hat{I}}}{x_{\hat{J}}} \right)^{2\gamma}(-x_{\hat{J}})^{2\gamma}\right)\\
&\hspace{10mm}\times\left(\prod_{n+1 \leq \tilde{I} <\tilde{J} \leq N}
     \left( 1-\frac{x_{\tilde{I}}}{x_{\tilde{J}}} \right)^{2\gamma}(-x_{\tilde{J}})^{2\gamma}\right)
\left(\prod_{1 \leq \check{I} < \check{J} \leq n}
    | x_{\check{I}} - x_{\check{J}} |^{2\gamma}\right)\\
\Phi_N(\beta_1,\beta_2,\gamma;x):&=\Phi_N^{(N)}(\beta_1,\beta_2,\gamma;x).
\end{split}
\end{align}
The reason why we employ the complicated term $(1-\frac{x_I}{x_J})^{2\gamma}(-x_J)^{2\gamma}$ instead of $(x_I-x_J)^{2\gamma}$ is that 
    this term has following good properties:
\begin{align}
\left( 1-\frac{x}{y} \right)^{2\gamma}(-y)^{2\gamma}
&=\ex ^{-2\gamma\pi\im }
    \left( 1-\frac{y}{x} \right)^{2\gamma}(-x)^{2\gamma}
        \hspace{5mm}(2\pi > \mathrm{Arg} \: x > \mathrm{Arg} \: y \ge 0)\label{eq:prop_of_xj-xi_1}\\
\left( 1-\frac{x}{y} \right)^{2\gamma}(-y)^{2\gamma}
&=(x-y)^{2\gamma}.
    \hspace{5mm} (x \in \mathbb{R}^+)\label{eq:prop_of_xj-xi_2}
\end{align}

Let $f(x)=f(x_1,x_2,\cdots,x_N)$ be a holomorphic function in the region excluding $x \in (1,+\infty)$
    and be invariant under the permutations of $x_1,x_2,\cdots,x_N$.
For the sake of convenience, we introduce the symbol $\mathrm{\bf{s}}(z)$ that stands for
\begin{align}
\mathrm{\bf{s}}(z):&= \sin \pi z.
\end{align}

%%%%%%%%%%%%%%%%%%%%%%%%%%%%%%%%%%%% figure 2 %%%%%%%%%%%%%%%%%%%%%%%%%%%%%%%%%%%%%%%%%%%%%%%%%%%%%%
%%%%%%%%%%%%%%%%%%%%%%%%%%%%%%%%%%%%%%%%%%%%%%%%%%%%%%%%%%%%%%%%%%%%%%%%%%%%%%%%%%%%%%%%%%%%%%%%%%%%%%%%%

%%%%%%%%%%%%%%%%%%%%%%%%%%%%%%%%%%% figure 3 %%%%%%%%%%%%%%%%%%%%%%%%%%%%%%%%%%%%%%%%%%%%%%%%%%%%%%%%%
%%%%%%%%%%%%%%%%%%%%%%%%%%%%%%%%%%%%%%%%%%%%%%%%%%%%%%%%%%%%%%%%%%%%%%%%%%%%%%%%%%%%%%%%%%%%%%%%%%%%%%

Secondly, we show the following equations:
\begin{align} 
\left(\prod_{I=1}^N\int_{\tilde{\mathcal{C}}_\rho}\de x_I \right)
    \Phi_N^{(0)}(x)f(x)
&=\left(
    \prod_{I=1}^N\ex^{\im\pi\gamma (I-N)}\frac{\mathrm{\bf{s}}(\gamma I)}{\mathrm{\bf{s}}( \gamma) }
\right)
    \int_{1+0\im\preceq x_1\preceq \cdots\preceq x_N\preceq 1+0-0\im}
        \left(\prod_{I=1}^N\de x_I \right)\Phi_N^{(0)}(x)f(x)
\label{eq:exchange_constant}\\
\left(\prod_{I=1}^N\int_{\mathcal{C}[0,1]}\de x_I \right)
    \Phi_N^{(0)}(x)f(x)
&=\left(
    \prod_{I=1}^N\ex^{\im\pi\gamma (I-N)}\frac{\mathrm{\bf{s}}(\gamma I)}{\mathrm{\bf{s}}( \gamma) }
\right)
    \int_{1+0\im\succeq  x_1\succeq \cdots\succeq x_N\succeq 1-0-0\im}
        \left(\prod_{I=1}^N\de x_I \right)\Phi_N^{(0)}(x)f(x).
\label{eq:exchange_constant2}
\end{align}
Let $\mathcal{S}_N$ be the symmetric group of degree $N$. 
We define the action of $\sigma \in \mathcal{S}_N$ on a symmetric function $g(x)$ as follows:
\begin{align}
\sigma\cdot g(x_1,x_2,\cdots,x_N) = g(x_{\sigma(1)},x_{\sigma(2)},\cdots,x_{\sigma(N)}).
\end{align}
The equation (\ref{eq:exchange_constant}) can be written as
\begin{align}
\begin{split}
\sum_{\sigma\in\mathcal{S}_N}&\int_{1+0\im\preceq x_1\preceq \cdots\preceq x_N\preceq 1+0-0\im}\left(\prod_{I=1}^N\de x_I \right) \sigma \cdot \Phi_N^{(0)}(x) f(x)\\
&\hspace{5mm}=\left(
    \prod_{I=1}^{N}\ex^{\im\pi\gamma (I-N)}\frac{\mathrm{\bf{s}}(\gamma I)}{\mathrm{\bf{s}}( \gamma) }
\right)
\int_{1+0\im\preceq x_1\preceq \cdots\preceq x_N\preceq 1+0-0\im}\left(\prod_{I=1}^N\de x_I \right)\Phi_N^{(0)}(x)f(x).
\end{split}
\end{align}
Let $\rho_{i,j}$ be a transposition and $\varpi_{n}$ be a cyclic permutation defined as follows:
\begin{align}
\rho_{i,j}(k)&=\left\{
  \begin{array}{ccl}
    j   & \  &(k=i)  \\
    i   & \  &(k=j)  \\
    k   & \  &(k\neq i,j)  \\
  \end{array}
\right.\\
\varpi_n(k)&=\left\{
  \begin{array}{ccl}
    k   & \  &(k<n)  \\
    N+1   & \  &(k=n)  \\
    k-1   & \  &(k>n)  \\
  \end{array}
\right.
.
\end{align}
Note that 
\begin{align}
\varpi_n=\rho_{N+1,N}\cdot \rho_{N,N-1}\cdot \cdots \cdot \rho_{n+1,n}.
\end{align}
The eq.(\ref{eq:prop_of_xj-xi_1}) implies
\begin{align}
\rho_{i,i+1} \cdot \Phi_N^{(0)}(x_1,x_2,\cdots,x_N) = \ex^{-2 \pi \gamma \im} \Phi_N^{(0)}(x_1,x_2,\cdots,x_N)
\hspace{5mm}(1+0\im\preceq x_1\preceq \cdots\preceq x_N\preceq 1+0-0\im).
\end{align}
Furthermore, introduce
\begin{align}
\mathcal{S}_{N+1}^N:&=\{\sigma\in\mathcal{S}_{N+1}|\sigma(N+1)=N+1\}.
\end{align}
Note that 
\begin{align}
\mathcal{S}_{N+1} &= \bigcup_{n}  \left(\mathcal{S}_{N+1}^N \cdot \varpi_n\right)\\
\emptyset &= \left(\mathcal{S}_{N+1}^N \cdot \varpi_i \right) \cap \left(\mathcal{S}_{N+1}^N \cdot \varpi_j\right)\hspace{5mm}(i \neq j),
\end{align}
where
\begin{align}
\mathcal{S}_{N+1}^N \cdot \varpi_n=\{ \sigma \cdot \varpi_n \in\mathcal{S}_{N+1}|\:\sigma\in\mathcal{S}_{N+1}^{N} \}.
\end{align}
Now, we employ the mathematical induction on $N$ to show (\ref{eq:exchange_constant}).
If $N=2$, (\ref{eq:exchange_constant}) is obvious.
Suppose that if $N=k$, (\ref{eq:exchange_constant}) is true.
For $N=k+1$, we obtain
\begin{align}
\begin{split}
&\sum_{\sigma\in\mathcal{S}_{k+1}}
    \int_{1+0\im\preceq x_1\preceq \cdots\preceq x_{k+1}\preceq 1+0-0\im}
        \left(\prod_{I=1}^{k+1}\de x_I \right)
            \sigma \cdot\Phi_{k+1}^{(0)}(x)f(x)\\
&\hspace{10mm}=\sum_{n=1}^{k+1}\sum_{\sigma_k\in\mathcal{S}^k_{k+1}}
    \int_{1+0\im\preceq x_1\preceq \cdots\preceq x_{k+1}\preceq 1+0-0\im}
        \left(
            \prod_{I=1}^{k+1}\de x_I
        \right)
            \sigma_k \cdot\varpi_n \cdot\Phi_{k+1}^{(0)}(x)f(x)\\
&\hspace{10mm}=\sum_{n=1}^{k+1}\ex^{-2 \pi \im \gamma (n-1)}\sum_{\sigma_k\in\mathcal{S}^k_{k+1}}
    \int_{1+0\im\preceq x_1\preceq \cdots\preceq x_{k+1}\preceq 1+0-0\im}
        \left(
            \prod_{I=1}^{k+1}\de x_I
        \right)
            \sigma_k \cdot \Phi_{k+1}^{(0)}(x)f(x)\\
&\hspace{10mm}=
    \left(
        \prod_{I=1}^{k+1}\ex^{\im\pi\gamma (I-k)}\frac{\mathrm{\bf{s}}(\gamma I)}{\mathrm{\bf{s}}( \gamma) }
    \right)
    \int_{1+0\im\preceq x_1\preceq \cdots\preceq x_{k+1}\preceq 1+0-0\im}
    \left(\prod_{I=1}^{k+1}\de x_I \right) \Phi_{k+1}^{(0)}(x)f(x).
\end{split}
\end{align}
Thus, (\ref{eq:exchange_constant}) is proven.
The eq. (\ref{eq:exchange_constant2}) is shown by the same proof.

Thirdly, we show 
\begin{align}
\left(\prod_{I=1}^N\int_0^1\de x_I \right)
    \Phi_N^{(N)}(x)f(x)
=\left(\frac{\im}{2}\right)^N\left(
    \prod_{I=1}^N\frac{\ex^{\im\pi\gamma (I-N)}}{\mathrm{\bf{s}}(\beta+(N-I) \gamma) }
\right)
        \left(\prod_{I=1}^N\int_{\mathcal{C}_\rho}\de x_I \right)\Phi_N^{(0)}(x)f(x).
\label{eq:real_to_Complex}
\end{align}
Now, we consider the case 
\begin{align}
0 &\le x_1 \le  x_2 \cdots \le x_{n} \le 1\label{eq:x<x}\\
1+0\im &\preceq x_{n+2} \preceq \cdots \preceq x_{N} \preceq 1+0-0\im\label{eq:0<x<1<x<}
\end{align}
The $n$ points $x_1,\cdots,x_n$ lie on the segments $\mathcal{C}_{[0,1]}$.
Let us consider
\begin{align}
\int_{\mathcal{C}_{\rho}} \Phi^{(n)}_N(\beta_1,\beta_2,\gamma;x) f(x) \de x_{n+1}.
\end{align}
This integral vanishes as there is no singularity in the region enclosed by this contour.
Converting the integral over $\mathcal{C}_{[0,1]}$ into that over the segment $(0,1)$ on the real axis, we obtain
\begin{align}
\begin{split}
0
&=
    -\Bigg[ e^{-\im\pi\beta_1}\left(
        \int_{0}^{x_1}
        +e^{-2\im\pi \gamma}\int_{x_1}^{x_2}
        +e^{-4\im\pi \gamma}\int_{x_2}^{x_3}
        +\cdots
        +e^{-2n\im\pi \gamma} \int_{x_{n}}^{1}
    \right)\\
&\hspace{25mm}
    - e^{\im\pi\beta_1}\left(
        \int_{0}^{x_1}
        +e^{2\im\pi \gamma}\int_{x_1}^{x_2}
        +e^{4\im\pi \gamma}\int_{x_2}^{x_3}
        +\cdots
        +e^{2n\im\pi \gamma} \int_{x_{n}}^{1}
    \right)\Bigg]\\
&\hspace{25mm}
    \times \Phi_N^{(n+1)}(\beta_1,\beta_2,\gamma;x) f(x) \de x_{n+1}\\
&\hspace{15mm}+
    \left(
        \int_{1+0\im \preceq x_{n+1} \preceq x_{n+2}}
        +\cdots
        +\int_{x_{N}\preceq x_{n+1} \preceq 1-0\im}
    \right)
\Phi^{(n)}_N(\beta_1,\beta_2,\gamma;x) f(x) \de x_{n+1}.
\end{split}
\end{align}
Here, the phase factors appear due to the replacement $\Phi_N^{(n)}\to\Phi_N^{(n+1)}$.
\par
Let us integrate the above expression over $x_1,\cdots,x_{n},x_{n+2},\cdots,x_{N}$,
    keeping the ordering (\ref{eq:x<x}) and (\ref{eq:0<x<1<x<}).
In this formula, the integrals over the real axis can be converted into those over the region
    $0\le  x_1\le  \cdots \le x_{n+1}\le 1$ and $1+0\im \preceq x_{n+2} \preceq \cdots \preceq x_N \preceq 1+0-0\im$
    by the appropriate interchange of $x_I$'s.
The remaining integrals can also be converted into those over the region  
    $0\le  x_1\le  \cdots \le x_{n}\le 1$ and $1+0\im \preceq x_{n+1} \preceq \cdots \preceq x_N \preceq 1+0-0\im$.
We obtain
\begin{align}
\begin{split}\label{eq:gen_Iguri1}
    -\frac{2\im\mathrm{\bf{s}}((n+1)\gamma) \mathrm{\bf{s}} (\beta_1 + n\gamma)}{\mathrm{\bf{s}} (\gamma)}
&
    \int_{
        \begin{subarray}{c}
            0 \le x_1 \le \cdots \le x_{n+1} \le 1\\
            1+0\im \preceq x_{n+2} \preceq \cdots \preceq x_{N} \preceq 1-0\im
        \end{subarray}
    }
    \left(\prod_{I=1}^{N}  \de x_I\right)
    \Phi_N^{(n+1)}(\beta_1,\beta_2,\gamma;x) f(x) \\
=&
    \int_{
        \begin{subarray}{c}
            0 \le x_1 \le \cdots \le x_{n} \le 1\\
            1+0\im \preceq x_{n+1} \preceq \cdots \preceq x_{N} \preceq 1-0\im
        \end{subarray}
    }
    \left(\prod_{I=1}^{N}  \de x_I\right)
    \Phi_N^{(n)}(\beta_1,\beta_2,\gamma;x) f(x),
\end{split}
\end{align}
where we use the following formula:
\begin{align}
\sum_{j=0}^n \mathrm{\bf{s}} (\beta_1 + 2 j \gamma )&=\frac{\mathrm{\bf{s}}((n+1)\gamma) \mathrm{\bf{s}} (\beta_1 + n\gamma)}{\mathrm{\bf{s}} (\gamma)}.
\end{align}
Using (\ref{eq:gen_Iguri1}) repeatedly and (\ref{eq:exchange_constant}), we obtain (\ref{eq:real_to_Complex}).

Fourthly, we show
\begin{align}
\left(\prod_{I=1}^N\int_0^1\de x_I \right)
    \Phi_N^{(N)}(x)f(x)
=\left(-\frac{\im}{2}\right)^N\left(
    \prod_{I=1}^N\frac{\ex^{\im\pi\gamma (I-N)}}{\mathrm{\bf{s}}(\beta+(N-I) \gamma) }
\right)
        \left(\prod_{I=1}^N\int_{\mathcal{C}_{[0,1]}}\de x_I \right)\Phi_N^{(0)}(x)f(x).
\label{eq:real_to_segment}
\end{align}
Consider
\begin{align}
\int_{\mathcal{C}_{\rho}} \Phi^{(0)}_N(\beta_1,\beta_2,\gamma;x) f(x) \de x_{n+1}.
\end{align}
with the ordering  (\ref{eq:0<x<1<x<}) and
\begin{align}
1+0\im\succeq  x_1\succeq \cdots\succeq x_N\succeq 1-0-0\im.
\end{align}
We follow the proof of (\ref{eq:real_to_Complex})
    but this time not converting the integral over $\mathcal{C}_{[0,1]}$ into that over the segment $(0,1)$ on the real axis.
By using (\ref{eq:exchange_constant}) and (\ref{eq:exchange_constant2}),
we obtain 
\begin{align}
\begin{split}
\left(\prod_{I=1}^N\int_{\tilde{\mathcal{C}}_\rho}\de x_I \right)\Phi_N^{(0)}(x)f(x)=(-1)^N
        \left(\prod_{I=1}^N\int_{\mathcal{C}_{[0,1]}}\de x_I \right)\Phi_N^{(0)}(x)f(x).
\end{split}
\end{align}
Now, we derive (\ref{eq:real_to_segment}) from (\ref{eq:real_to_Complex}).

Fifthly, we prove
\begin{align}
\begin{split}\label{eq:gen_Iguri3}
&\left(\prod_{I=1}^{N} \int_0^1 \de x_I\right)
    \Phi_N(\beta_1,\beta_2,\gamma;x) f(x)
=
\left(\frac{\im}{2}\right)^N\left(
    \prod_{I=1}^N\frac{\ex^{\im\pi\gamma (I-N)}}{\mathrm{\bf{s}}(\beta+(N-I) \gamma) }
\right)\\
    &\hspace{32mm}\times\left(\prod_{I=1}^{N} \int_{\tilde{\mathcal{D}}(1/\rho,1)} \de x_{I}\right)
    \Phi^{(0)}_N(1-\beta_1-\beta_2 -2(N-1)\gamma,\beta_2,\gamma;x)f\Big(\frac{1}{x}\Big).
\end{split}
\end{align}
By the transformation $x \to 1/x$,
    we obtain
\begin{align}
\begin{split}
&\left(\prod_{I=1}^{N} \int_{\tilde{\mathcal{C}}_\rho} \de x_{I}\right)
    \Phi_N(\beta_1,\beta_2,\gamma;x) f(x)\\
&\hspace{25mm}=
    \left(\prod_{I=1}^{N} \int_{\tilde{\mathcal{D}}(1/\rho,1)} \de x_{I}\right)
    \Phi_N(1-\beta_1-\beta_2 -2(N-1)\gamma,\beta_2,\gamma;x)f\Big(\frac{1}{x}\Big),\\
\end{split}
\end{align}
where $f\Big(\frac{1}{x}\Big)$ is obtained by replacing $x_I$ by $1/x_I$ in $f(x)$
and $\tilde{\mathcal{D}}_{1/\rho,z}$ is defined by Fig.\ref{fig:CD}.
In obtaining the above equation, we have used (\ref{eq:prop_of_xj-xi_2}) with $x=1$ and convert the variable $x_I$ to $x_{N+1-I}$.
By using (\ref{eq:real_to_Complex}), we have (\ref{eq:gen_Iguri3}).
Notice that this formula holds independently of the radius parameter $\rho$.

Let us consider the limit of (\ref{eq:gen_Iguri3}) as $1/\rho \to 0$. 
Then, $\tilde{\mathcal{D}}(1/\rho,1)$ reduce to $-\mathcal{C}_{[0,1]}$.
Suppose that $f(1/x)$ is holomorphic function at $x \in (0,1)$.
Then, we can apply (\ref{eq:real_to_segment}) to (\ref{eq:gen_Iguri3}) and obtain
\begin{align}
\begin{split}
&\left(\prod_{I=1}^{N} \int_0^1 \de x_I\right)
    \Phi_N(\beta_1,\beta_2,\gamma;x) f(x)\\
&\hspace{15mm}=C_N(\beta_1,\beta_2,\gamma)\left(\prod_{I=1}^{N} \int_0^1 \de x_{I}\right)
    \Phi^{(0)}_N(1-\beta_1-\beta_2 -2(N-1)\gamma,\beta_2,\gamma;x)f\Big(\frac{1}{x}\Big),
\end{split}
\end{align}
where
\begin{align}
C_N(\beta_1,\beta_2,\gamma)&:=
\prod_{I=1}^N\frac{\mathrm{\bf{s}}(\beta_1+\beta_2+(2N-I-1)\gamma)}{\mathrm{\bf{s}}(\beta_1+(N-I)\gamma)}.
\end{align}
The above formula has been shown by Iguri \cite{I}.

%%%%%%%%%%%%%%%%%%%%%%%%%%%%%%%%%%%%%%%%%%%%%%%%%%%%%%%%%%%%%%%%%%%%%%%%%%%%%%%%%%%%%%%%%


\begin{thebibliography}{99}

\bibitem{SW9408}
N. Seiberg and E. Witten,
``Monopoles, duality and chiral symmetry breaking
in $N=2$ supersymmetric QCD,''
Nucl. Phys. B \textbf{431} (1994) 484-550,
[arXiv:hep-th/9408099].

\bibitem{HO}
A.~Hanany and Y.~Oz,
``On the Quantum Moduli Space of Vacua of 
$N=2$ Supersymmetric $SU(N_c)$ Gauge Theories,''
Nucl.\ Phys.\  B {\bf 452}, 283-312 (1995)
[arXiv:hep-th/9505075].

\bibitem{APS}
P.C. Argyres, M.R. Plesser and A. Shapere,
``The Coulomb Phase of $N=2$ Supersymmetric QCD,''
Phys. Rev. Lett. \textbf{75} (1995) 1699-1702,
[arXiv:hep-th/9505100].

\bibitem{Oht1}
Y.~Ohta,
``Prepotential of $N = 2$ $SU(2)$ Yang-Mills gauge theory coupled with a  massive
matter multiplet,''
J.\ Math.\ Phys.\  {\bf 37}, 6074-6085 (1996)
[arXiv:hep-th/9604051].

\bibitem{Oht2}
Y.~Ohta,
``Prepotentials of $N = 2$ $SU(2)$ Yang-Mills theories coupled with massive
matter multiplets,''
J.\ Math.\ Phys.\  {\bf 38}, 682-696 (1997)
[arXiv:hep-th/9604059].

\bibitem{DHKP9609}
E.~D'Hoker, I.~M.~Krichever and D.~H.~Phong,
``The effective prepotential of $N = 2$ supersymmetric $SU(N_c)$ gauge
theories,''
Nucl.\ Phys.\  B {\bf 489}, 179-210 (1997)
[arXiv:hep-th/9609041].

\bibitem{MasSuz}
T.~Masuda and H.~Suzuki,
``Periods and prepotential of $N = 2$ $SU(2)$ supersymmetric Yang-Mills  theory
with massive hypermultiplets,''
Int.\ J.\ Mod.\ Phys.\  A {\bf 12}, 3413-3431 (1997)
[arXiv:hep-th/9609066].

\bibitem{DHKP9610}
E.~D'Hoker, I.~M.~Krichever and D.~H.~Phong,
``The renormalization group equation in $N = 2$ supersymmetric gauge
theories,''
Nucl.\ Phys.\  B {\bf 494}, 89-104 (1997)
[arXiv:hep-th/9610156].

\bibitem{DKM9611}
N.~Dorey, V.~V.~Khoze and M.~P.~Mattis,
``On $N = 2$ supersymmetric {QCD} with 4 flavors,''
Nucl.\ Phys.\  B {\bf 492}, 607-622 (1997)
[arXiv:hep-th/9611016].


%%%%%%%%%%%%%%%%%%%%%%%%%%%%%%%%%%%%%%%%%%%%%%%%%%%%%%%%%%%%%%%%

\bibitem{AGT}
  L.~F.~Alday, D.~Gaiotto and Y.~Tachikawa,
  ``Liouville Correlation Functions from Four-dimensional Gauge Theories,''
Lett.\ Math.\ Phys.\  {\bf 91}, 167-197 (2010)
  [arXiv:0906.3219 [hep-th]].

\bibitem{Wyllard}
  N.~Wyllard,
  ``$A_{N-1}$ conformal Toda field theory correlation functions from conformal
  $\mathcal{N}=2$ $SU(N)$ quiver gauge theories,''
  JHEP \textbf{0911}, 002 (2009)
  [arXiv:0907.2189 [hep-th]].


\bibitem{MMM0907}
A.~Marshakov, A.~Mironov and A.~Morozov,
``On Combinatorial Expansions of Conformal Blocks,''
arXiv:0907.3946 [hep-th].

\bibitem{MM0908a}
A.~Mironov and A.~Morozov,
``The Power of Nekrasov Functions,''
Phys. Lett. B \textbf{680}, 188-194 (2009)
[arXiv:0908.2190 [hep-th]].

\bibitem{MM0908b}
A.~Mironov and A.~Morozov,
``On AGT relation in the case of $U(3)$,''
Nucl. Phys. B \textbf{825}, 1-37 (2010)
[arXiv:0908.2569 [hep-th]].

\bibitem{FatLit}
  V.~A.~Fateev and A.~V.~Litvinov,
  ``On AGT conjecture,''
  JHEP {\bf 1002}, 014 (2010)
  [arXiv:0912.0504 [hep-th]].

\bibitem{HJS}
L.~Hadasz, Z.~Jask\'{o}lski and P.~Suchanek,
  ``Proving the AGT relation for $N_f = 0,1,2$ antifundamentals,''
  JHEP {\bf 1006}, 046 (2010)
  [arXiv:1004.1841 [hep-th]].


%%%%%%%%%%%%%%%%%%%%%%%%%%%%%%%%%%%%%%%%%%%%%%%%%%%%%%%%


\bibitem{gai0908}
D.~Gaiotto,
``Asymptotically free $\mathcal{N}=2$ theories
and irregular conformal blocks,''
arXiv:0908.0307 [hep-th].

\bibitem{MMM0909a}
A.~Marshakov, A.~Mironov and A.~Morozov,
``On non-conformal limit of the AGT relations.,''
Phys. Lett. B \textbf{682}, 125-129 (2009)
[arXiv:0909.2052 [hep-th]].

\bibitem{MMM0909b}
A.~Marshakov, A.~Mironov and A.~Morozov,
``Zamolodchikov asymptotic formula and instanton expansion
in $\mathcal{N}=2$ SUSY $N_f = 2 N_c$ QCD,''
JHEP \textbf{11}, 048 (2009)
[arXiv:0909.3338 [hep-th]].

\bibitem{AM0910}
V.~Alba and And.~Morozov,
``Non-conformal limit of AGT relation from the 1-point torus conformal block,''
JETP Lett. \textbf{90}, 708-712 (2009)
[arXiv:0911.0363 [hep-th]].

%%%%%%%%%%%%%%%%%%%%%%%%%%%%%%%%%%%%%%%%%%%%%%%

\bibitem{DV}
  R.~Dijkgraaf and C.~Vafa,
  ``Toda Theories, Matrix Models, Topological Strings, and $N=2$ Gauge Systems,''
  arXiv:0909.2453 [hep-th].

\bibitem{IMO}
  H.~Itoyama, K.~Maruyoshi and T.~Oota,
  ``The Quiver Matrix Model and 2d-4d Conformal Connection,''
Prog. Theor. Phys. \textbf{123}, 957-987 (2010)
[arXiv:0911.4244 [hep-th]].

%%%%%%%%%%%%%%%%%%%%%%%%%%%%%%%%%%%%%%%%%%%%%%%%%%%
\bibitem{Witten}
  E.~Witten,
  ``Solutions of four-dimensional field theories via $M$-theory,''
  Nucl.\ Phys.\  B {\bf 500}, 3-42 (1997)
  [arXiv:hep-th/9703166].

\bibitem{Gaiotto}
  D.~Gaiotto,
  ``$N=2$ dualities,''
  arXiv:0904.2715 [hep-th].

%%%%%%%%%%%%%%%%%%%%%%%%%%%%%%%%%%%%%%%%%%%%


\bibitem{EM}
 T.~Eguchi and K.~Maruyoshi,
  ``Penner Type Matrix Model and Seiberg-Witten Theory,''
  JHEP {\bf 1002}, 022 (2010)
  [arXiv:0911.4797 [hep-th]]; \\
T.~Eguchi and K.~Maruyoshi,
``Seiberg-Witten theory, matrix model and AGT relation,''
arXiv:1006.0828 [hep-th].

%%%%%%%%%%%%%%%%%%%%%%%%%%%%%%%%%%%%%%%%%%%%%%%%%%%

\bibitem{DF}
  V.~S.~Dotsenko and V.~A.~Fateev,
  ``Conformal algebra and multipoint correlation functions in  2D statistical
  models,''
  Nucl.\ Phys.\  B {\bf 240}, 312-348 (1984); \\
V.~S.~Dotsenko and V.~A.~Fateev,
``Four Point Correlation Functions And The Operator Algebra In The
Two-Dimensional Conformal Invariant Theories With The Central Charge $c \leq 1$,''
Nucl.\ Phys.\  B {\bf 251}, 691-734 (1985).

\bibitem{MMS0911}
  A.~Mironov, A.~Morozov and Sh.~Shakirov,
  ``Matrix Model Conjecture for Exact BS Periods and Nekrasov Functions,''
  JHEP {\bf 1002}, 030 (2010)
  [arXiv:0911.5721 [hep-th]].

\bibitem{MMS1001}
  A.~Mironov, A.~Morozov and Sh.~Shakirov,
  ``Conformal blocks as Dotsenko-Fateev Integral Discriminants,''
  arXiv:1001.0563 [hep-th].

\bibitem{IO5}
H.~Itoyama and T.~Oota,
``Method of Generating $q$-Expansion Coefficients for Conformal Block and 
$\mathcal{N}=2$ Nekrasov Function by $\beta$-Deformed Matrix Model,''
  Nucl.\ Phys.\  B {\bf 838}, 298-330 (2010)
  [arXiv:1003.2929 [hep-th]].

\bibitem{BT0909}
G.~Bonelli and A.~Tanzini,
``Hitchin systems, $\mathcal{N}=2$ gauge theories and W-gravity,''
Phys.\ Lett.\  B {\bf 691}, 111-115 (2010)
[arXiv:0909.4031 [hep-th]].


\bibitem{MMM1003}
  A.~Mironov, Al.~Morozov and And.~Morozov,
  ``Matrix model version of AGT conjecture and generalized Selberg integrals,''
  arXiv:1003.5752 [hep-th].

\bibitem{KPW}
C.~Koz\c{c}az, S.~Pasquetti and N.~Wyllard,
``A \& B model approaches to surface operators and Toda theories,''
arXiv:1004.2025 [hep-th].

\bibitem{MS1004}
  A.~Morozov and S.~Shakirov,
  ``The matrix model version of AGT conjecture and CIV-DV prepotential,''
  arXiv:1004.2917 [hep-th].

\bibitem{AY1004}
  H.~Awata and Y.~Yamada,
  ``Five-dimensional AGT Relation and the Deformed $\beta$-ensemble,''
  arXiv:1004.5122 [hep-th].

\bibitem{NX1005}
D.~Nanopoulos and D.~Xie,
``Hitchin Equation, Irregular Singularity, and $N=2$ Asymptotical Free
Theories,''
arXiv:1005.1350 [hep-th].

\bibitem{Tai1006}
T.~S.~Tai,
``Triality in $SU(2)$ Seiberg-Witten theory and Gauss hypergeometric
  function,''
  arXiv:1006.0471 [hep-th].

\bibitem{NX1006}
D.~Nanopoulos and D.~Xie,
``$N=2$ Generalized Superconformal Quiver Gauge Theory,''
arXiv:1006.3486 [hep-th].

\bibitem{KMS1007}
  S.~Kanno, Y.~Matsuo and S.~Shiba,
  ``Analysis of correlation functions in Toda theory and AGT-W relation for
  $SU(3)$ quiver,''
  arXiv:1007.0601 [hep-th].


%%%%%%%%%%%%%%%%%%%%%%%%%%%%%%%%%%%%%%%%%%%%%%%%%

\bibitem{sel}
A.~Selberg,
``Bemerkninger om et multipelt integral,''
Norsk Mat.\ Tidsskr.\ \textbf{26}, 71-78 (1944).

\bibitem{mac1}
I.~G.~Macdonald,
``Commuting differential operators and zonal spherical
functions,''
in \textit{Algebraic Groups Utrecht 1986},
Proceedings of a Symposium in Honour of T.~A.~Springer,
Lecture Notes in Math. \textbf{1271}, 189-200, 
ed. by A.~M.~Cohen, W.~H.~Hesselink, W.~L.~J.~van der Kallen and J.~R.~Strooker,
Springer (1987).

\bibitem{kad}
K.~W.~J.~Kadell,
``The Selberg-Jack symmetric functions,''
Adv. Math. \textbf{130}, 33-102 (1997).

\bibitem{kan}
J.~Kaneko,
``Selberg integrals and hypergeometric functions
associated with Jack polynomials,''
SIAM.\ J.\ Math.\ Anal.\ \textbf{24}, 1086-1110 (1993).

\bibitem{I}
S.~Iguri,
``On a Selberg-Schur integral,''
Lett. Math. Phys. \textbf{89}, 141-158 (2009) \\ {}
[arXiv:0810.5552 [math-ph]].


\end{thebibliography}
\end{document}